\documentstyle[12pt,epsf]{article}

\def \epsf#1#2{\epsfysize=#2\epsfbox{#1}}
\def\epsfx#1#2{\epsfxsize=#2\epsfbox{#1}}

\def\centered{\centering\leavevmode}

\catcode`\@=11
\def\indentation{\@afterindenttrue}
\catcode`\@=12
\def\isection#1{\section{#1}\indentation}
\def\isubsection#1{\subsection{#1}\indentation}

\def\mc{\multicolumn{1}{c|}{}}

\def\${\ifmmode\end{equation}\else\begin{equation}\fi}

\def\be{\begin{eqnarray}}
\def\ee{\end{eqnarray}}
\def\no{\nonumber\\}

\def\ble{\begingroup\interdisplaylinepenalty=10000\be}
\def\ele{\ee\endgroup}

\def\(#1){(\ref{#1})}
\def\_#1_{\qquad\hbox{#1}\qquad}

\def\bvec#1{\Biggl({#1}\Biggr)} 
\def\cvec#1{\left(\begin{array}{c}#1\\ \end{array}\right)}
\def\matr#1{\left(\begin{array}{cccc}#1\\ \end{array}\right)}

\def\mm{\mskip-\medmuskip}
\def\cvcc#1{\left(\mm\begin{array}{c}#1\\ \end{array}\mm\right)}
\def\cvcC#1{\left(\mm\mm\begin{array}{c}#1\\ \end{array}\mm\mm\right)}

\def\0{} 

\def\wk{{\w k \over E \sqrt{p^2+k^2}}}
\def\pM{{ p M \over E \sqrt{p^2+k^2}}}
\def\xwk{{\w k_M \over E \sqrt{p^2+k_M^2}}}
\def\xpM{{ p M \over E \sqrt{p^2+k_M^2}}}

\def\overalign#1#2#3#4{\vcenter{\halign{%
       \hfil$##$&$##$\hfil\cr #1&#2\cr
       \noalign{\vskip6\fontdimen8\textfont3 \hrule
       \vskip3\fontdimen8\textfont3} #3&#4\cr}}}

\def\qsl{{\lower0.7pt\hbox{$\scriptscriptstyle L$}}}
\def\qsr{{\lower0.7pt\hbox{$\scriptscriptstyle R$}}}
\def\sl{{\mkern-1.5mu\qsl}}
\def\sr{{\mkern-1.5mu\qsr}}

\def\phc{\;+\;\hbox{H. c.}}
\def\Dslash{D\llap{\raise1.2pt\hbox{$\slash$}\kern0.35ex}}
\def\dslash{\d\llap{\raise1.5pt\hbox{$\slash$}\kern0.05ex}}

\def\mtilde{\mkern3.8mu\widetilde{\hbox to1ex{\hss$m$\hss}}\mkern3.8mu}
\def\Wtilde{\mkern5.8mu\widetilde{\hbox to1ex{\hss$W$\hss}}\mkern5.8mu}

\def\gsqr{\sqrt{g^2 + g'^2}}

\def\up{\uparrow}
\def\down{\downarrow}

\def\sech{\mathop{\rm sech}\nolimits} 
\def\re{\mathop{\rm Re}\nolimits}
\def\sgn{\mathop{\rm sgn}\nolimits}

\def\eps{\varepsilon}
\let\oldchi=\chi
\def\chi{\raise2pt\hbox{$\oldchi$}}
\def\l{\lambda}
\def\w{\omega}
\def\d{\partial}
\def\L{{\cal L}}
\def\M{{\cal M}}
\def\Mf{\M_{\mkern-2mu f}}
\def\Z{{\cal Z}}
\def\W{{\cal W}}
\def\vo{v_0}
\def\k#1{{\kappa_{#1}}}

\def\vpv{{v'\over v}}
\def\vppv{{v''\over v}}

\def\exp#1{{\rm e}^{#1}}
\def\eipx{\exp{-ip\cdot x}}
\def\epipx{\exp{ip\cdot x}}
\def\eperp{\exp{-i\w t+ipx}}

\def\p#1{\phi_{#1}}
\def\psqr{\p1^2+\p2^2}
\def\Amu{A^\mu}
\def\Bmu{B^\mu}
\def\ppl{\phi^{+}}
\def\pmi{\phi^{-}}

\def\f#1/#2{{#1\over #2}}
\def\textf#1/#2{{\textstyle{#1\over #2}}}
\def\rt{\sqrt{2}}
\def\half{\scriptstyle\raise.3ex\hbox{$\scriptscriptstyle 1$}%
          \kern-.18em/\kern-.14em%
          \lower.25ex\hbox{$\scriptscriptstyle 2$}}

\def\ghost{\hphantom{-}}

\def\openabove#1{\vbox{\vskip3pt\hbox{$#1$}}}
\def\openbelow#1{\vtop{\hbox{$#1$}\vskip2pt}}

\def\via(#1){\buildrel\scriptstyle\(#1)\over\to}

\def\square#1#2{{\vcenter{\vbox{
  \hrule height .#2pt
  \hbox{\vrule width .#2pt height #1pt
        \kern #1pt
        \vrule width .#2pt}
  \hrule height .#2pt}}}}
\def\dbox{\mskip2mu\square64\mskip2mu}

\def\pinf{+\infty}
\def\minf{-\infty}

\begin{document}

\rightline{hep-ph/9412270}
\rightline{RU-94-59}
\rightline{December 7, 1994}
\baselineskip=18pt
\vskip0.5in
\begin{center}
{\bf\LARGE Scattering from a Domain Wall}
\vskip0.05in
{\bf\LARGE in a Spontaneously Broken Gauge Theory}
\vskip0.5in
{\large Glennys R. Farrar}%
\footnote{Research supported in part by NSF-PHY-91-21039}
\vskip0.05in
{\it Department of Physics and Astronomy \\
     Rutgers University, Piscataway, NJ 08855, USA}
\vskip0.35in
{\large John W. McIntosh, Jr.}
\vskip0.05in
{\it Physics Department \\
     Princeton University, Princeton, NJ 08544, USA}
\end{center}
\vskip0.5in

{\bf Abstract:}
We study the interaction of particles with a domain wall
at a symmetry-breaking phase transition
by perturbing about the domain wall solution.
We find the particulate excitations
appropriate near the domain wall
and relate them to the particles present far from the wall
in the uniform broken and unbroken phases.
For a quartic Higgs potential
we find analytic solutions to the equations of motion and
derive reflection and transmission coefficients.
We discover several bound states for particles near the wall.
Finally, we apply our results to the
electroweak phase transition in the standard~model.

\thispagestyle{empty}
\newpage
\addtocounter{page}{-1}

\tableofcontents
\thispagestyle{empty}
\newpage
\addtocounter{page}{-1}

\isection{Introduction}

During a first-order phase transition
in which a gauge symmetry is spontaneously broken,
a number of effects depend on the scattering of particles
from the domain wall between the phases of broken and unbroken symmetry.
For instance, at the electroweak phase transition,
a difference in the reflection coefficients of
particles carrying opposite quantum numbers
could play a role in producing the baryon asymmetry of the universe.
Analytic solutions for quark scattering
from domain walls of two different profiles
are already in the literature~\cite{fs,ayala},
but so far there has been no analogous treatment
for the bosons of the theory.
The study of boson scattering at a symmetry-breaking phase transition
is complicated by the fact that
the two phases have different particle contents,
as explained by the Higgs mechanism.
Addressing this difficulty is the main focus of the present work.

We begin by considering the simplest gauge theory
with spontaneous symmetry breaking,
the Abelian Higgs model.
To permit a static domain wall to exist,
we study the theory at its transition temperature.
We find the modes of particulate excitation
appropriate near the domain wall
and express the interaction of these modes with the wall
in terms of a one-dimensional scattering potential.
We relate these internal modes
to the particle modes present in the asymptotic broken and unbroken phases.
All these results are valid for a general Higgs potential.

In section~\ref{sec:scatter}
we specialize to a quartic Higgs potential and obtain
analytic solutions to the one-dimensional
scattering equations for the internal modes,
including solutions which describe bound states.
Using the connections between internal and asymptotic modes
derived in section~\ref{sec:connect}
we compute scattering probabilities for the asymptotic particle states.

Having completed our analysis of the Abelian Higgs model,
we proceed in section~\ref{sec:sm}
to study the electroweak phase transition in the standard model.
As a nonabelian gauge theory with partial symmetry breaking,
the standard model should be
representative of the entire class of spontaneously broken gauge theories.
We find that our results for the Abelian Higgs model
may be adapted without difficulty
to the case of the bosons in the standard model.
We complete our discussion of the standard model
by applying the methods of the previous sections
to the case of the fermions.
Specifically, we obtain internal and asymptotic modes,
scattering potentials, and connection matrices for fermion scattering.

\section{Internal and asymptotic modes}
\label{sec:connect}
\isubsection{The model}

The Abelian Higgs model contains two fields,
a complex scalar field~$\phi$ and a $U(1)$~gauge field~$\Amu$.
The Lagrangian is
\$ \L = - \f1/4 F_{\mu\nu} F^{\mu\nu}
        + (D_\mu\phi)^* (D^\mu\phi)
        - V(\rt\,|\phi|). \$
The covariant derivative
\$ D_\mu = \d_\mu + ieA_\mu \$
gives $\phi$ a charge~$+e$.
We place only a few conditions on the Higgs potential~$V$.
There must be a minimum at zero for the unbroken-symmetry phase,
and another minimum at some~$\vo$ for the broken phase.
For a stable domain wall to exist we need $V(\vo) = V(0)$;
for convenience we take the common value to be zero.

We expand the field~$\phi$ into real and imaginary parts as
\$ \phi = \f1/\rt(\p1+i\p2). \$
The Lagrangian becomes
\$ \L = - \f1/4 F_{\mu\nu} F^{\mu\nu}
        + \f1/2 (\d_\mu\p1 - eA_\mu\p2)^2
        + \f1/2 (\d_\mu\p2 + eA_\mu\p1)^2
        - V(\sqrt{\psqr}). \$
Minimizing the action gives equations of motion
for the three fields $\p1$, $\p2$ and~$\Amu$:
\be
\dbox\p1 & = & \ghost e\p2\d\cdot A + 2e\Amu\d_\mu\p2
    + e^2 A^2 \p1 - V'(\sqrt{\psqr}) {\p1\over\sqrt{\psqr}}
    \qquad \label{full-eom-p1} \\
\dbox\p2 & = & -      e\p1\d\cdot A - 2e\Amu\d_\mu\p1
    + e^2 A^2 \p2 - V'(\sqrt{\psqr}) {\p2\over\sqrt{\psqr}} \qquad \\
\d_\nu F^{\mu\nu} & = & \ghost e(\p1\d^\mu\p2 - \p2\d^\mu\p1)
    + e^2 A^\mu (\psqr). \label{full-eom-A}
\ee
These equations of motion have a stable domain wall solution
with~$\p1 = v(z)$ and $\p2 = \Amu = 0$.
We will find this solution and study small perturbations about it.
We write
\$ \p1 = v(z) + h \label{p1-expansion} \$
and take $h$, $\p2$ and~$\Amu$ to be perturbatively small.

The term of order zero in equation~\(full-eom-p1)
is the condition for the field configuration~$v(z)$ to be stable:
\$ v'' = V'(v). \label{eom-v} \$
We allow the prime symbol to have two meanings:
since $V$~is a function of~$v$ we take $V'$~to be~$dV/dv$;
in all other cases a prime indicates a derivative with respect to~$z$.
Integrating equation~\(eom-v) twice gives the domain wall solution
\$ z = \int_{\vo/2}^{v(z)} {d\tilde v\over\sqrt{2V(\tilde v)}}
   \label{vev-integral} \$
in which the unbroken phase is taken to be at~$z = \minf$.
We will also make use of the solutions $v=\vo$ and~$v=0$
which describe uniform broken and unbroken phases.

The first-order terms in equations~\(full-eom-p1)--\(full-eom-A) consist of
an equation for the field~$h$
\$ \dbox h = - V''(v) h \label{eom-h} \$
and the following coupled equations for $\p2$ and~$\Amu$:
\be
\dbox\p2 & = & -ev\,\d\cdot A - 2e\Amu\d_\mu v
    - {V'(v)\over v}\p2 \label{eom-p2} \\
\d_\nu F^{\mu\nu} & = & \ghost e(v\d^\mu\p2 - \p2\d^\mu v)
    + (ev)^2 \Amu. \label{eom-A}
\ee
If we take the divergence of equation~\(eom-A)
and apply the condition~\(eom-v) on the background field
we get equation~\(eom-p2) times a factor~$ev$.
Thus equation~\(eom-p2) is redundant as long as~$v$ is not zero.

\isubsection{Particle content far from the wall}

To obtain the first-order equations
appropriate to the phase of unbroken symmetry
we set $v=0$ in equations~\(eom-h)--\(eom-A).
Equation~\(eom-h) for~$h$ becomes
\$ (\dbox + m^2)h = 0, \label{unbroken-eom-h} \$
where the mass~$m$ is defined by
\$ m^2 = V''(0). \label{mass-charged-scalars} \$
Since $v$ is zero, equation~\(eom-p2) is not redundant;
it reduces to
\$ (\dbox + m^2)\p2 = 0. \label{unbroken-eom-p2} \$
Equations \(unbroken-eom-h) and~\(unbroken-eom-p2)
combine to give a single Klein-Gordon equation for the complex field~$\phi$:
\$ (\dbox + m^2)\phi = 0. \label{unbroken-eom-phi} \$
The final first-order equation~\(eom-A) simplifies to
\$ \d_\nu F^{\mu\nu} = 0. \label{unbroken-eom-A} \$
These perturbative equations tell us that the unbroken phase
contains a complex scalar field~$\phi$ of mass~$m$
and a massless vector boson~$\Amu$.
Our linearized equations of motion
describe only the interaction of particles with the background field~$v$,
not the interaction of particles with each other,
so we must refer back to the original Lagrangian to find out that
the scalar field~$\phi$ carries charge~$+e$.

We define the following normalized solutions
of equations \(unbroken-eom-phi) and~\(unbroken-eom-A).
Their meaning and use are discussed below.
\$ \begin{array}{|c|rcl|} \cline{1-4}
\ppl& \phi&=&  \openabove{\f1/2 \eipx}\\
\pmi& \phi&=&  \openabove{\f1/2 \epipx}\\
A_T&  A^\mu&=& \openabove{\eps^\mu \eipx}\\ \cline{1-4}
\end{array} \label{unbroken-solutions} \$
We choose our coordinate system so that
the four-vector~$p^\mu$ is~$(\w,p,0,k)$.
The energy~$\w$ is always positive,
but the momenta $p$ and~$k$ may have either sign.
The solutions~\(unbroken-solutions) are normalized so that their
time-averaged $z$-component of energy flux is~$\f1/2 \w k$.
In the case of the gauge boson solution~$A_T$
the normalization assumes that the polarization vector~$\eps^\mu$ satisfies
\$ \eps^2 = -1. \$
Although we use complex notation, the field~$\Amu$ is real-valued;
the operation of taking the real part is implied.
We subdivide $A_T$ into two polarizations $A_{T1}$ and~$A_{T2}$
with polarization vectors
\be
\eps_{T1} &=& (0,k,0,-p)/\sqrt{p^2+k^2} \\
\eps_{T2} &=& (0,0,1,0).
\ee

Using these standard solutions
we can describe a field configuration
with a single complex number.
For example,
the complex number~$A_{T1}$
represents the solution
\$ \Amu = A_{T1}\,\eps_{T1}^\mu \eipx. \$
When a standard solution has negative energy
we define the coefficient of the solution
to be the complex conjugate of the given complex number.
For example,
the complex number~$\pmi$ represents the solution
\$ \phi = (\pmi)^* \, \textf1/2 \epipx. \$

Next we consider the phase of broken symmetry.
Setting~$v=\vo$
we find that the first-order equations~\(eom-h)--\(eom-A) reduce to
\be
(\dbox + \mtilde^2)h &=& 0 \label{broken-eom-h} \\
\d_\nu F^{\mu\nu} &=& e\vo\,\d^\mu\p2 + (e\vo)^2 A^\mu, \label{broken-eom-A}
\ee
where
\$ \mtilde^2 = V''(\vo). \label{mass-higgs} \$
Defining the variable
\$ \Bmu = \Amu + \d^\mu \biggl({\p2\over ev}\biggr) \label{define-B} \$
which is invariant under perturbatively small gauge transformations
allows us to simplify equation~\(broken-eom-A) to
\$ \d_\nu F_B^{\mu\nu} = M^2 \Bmu, \label{broken-eom-B} \$
where the mass~$M$ is given by
\$ M^2 = (e\vo)^2. \label{mass-gauge-boson} \$
According to equations \(broken-eom-h) and~\(broken-eom-B)
the broken phase contains
a Higgs scalar~$h$ of mass~$\mtilde$
and a massive vector boson~$\Bmu$ of mass~$M$.

The normalized solutions of the broken-phase equations are as follows:
\$ \begin{array}{|c|rcl|} \cline{1-4}
h& h&=&     \openabove{\eipx}\\
B& B^\mu&=& \eps^\mu \eipx\\ \cline{1-4}
\end{array} \$
In addition to transverse modes $B_{T1}$ and~$B_{T2}$
with the polarizations $\eps_{T1}$ and~$\eps_{T2}$ given above,
the massive vector boson has a longitudinal mode~$B_L$ with polarization
\$ \eps_L = \biggl( { \w \over M\sqrt{\w^2-M^2} } \biggr)
   ({\w^2-M^2 \over \w}, p, 0, k). \$

\isubsection{Separation of scalar internal modes}

We will reduce the equations of motion~\(eom-h)--\(eom-A)
in the domain wall background
to independent scalar equations of the form
\$ [\dbox + U(z)]\,S = 0, \label{eom-scalar} \$
where $S$~is a scalar mode and $U$~is the potential it sees.
We take all fields to be eigenstates of energy and transverse momentum,
that is, to contain a factor~$\exp{-i\w t + ipx}$.
Once we define the positive quantity~$E$ by
\$ E^2 = \w^2-p^2, \$
equation~\(eom-scalar) becomes the Schr\"odinger equation
\$ [-\d_3^2 + U(z)]\,S = E^2 S \$
for a particle with nonrelativistic energy~$E^2$,
so our intuition for one-di\-men\-sion\-al scattering
can be applied to the potential~$U$.
Equation~\(eom-scalar) shows that the asymptotic values~$U(\pm\infty)$
are to be identified with
the mass squared of the scalar mode in the broken and unbroken phases.

Equation~\(eom-h) for~$h$ is already in the required scalar form---%
we need only define the potential
\$ U_h(z) = V''(v(z)). \$

To extract scalar modes from equation~\(eom-A) for~$\Amu$,
the first thing we do is write the equation
in terms of the variable~$\Bmu$~\(define-B):
\$ \d_\nu F_B^{\mu\nu} = (ev)^2 \Bmu. \label{eom-B} \$
We can get two scalar modes by making the ansatz~$B^3 = 0$.
In this case, taking the divergence of equation~\(eom-B)
produces the relation~$\d\cdot B = 0$,
which allows us to simplify equation~\(eom-B) to
\$ -\dbox \Bmu = (ev)^2 \Bmu. \$
Factoring out a constant polarization by writing
\$ \Bmu = \eps^\mu\tau \label{define-tau} \$
gives a scalar equation for~$\tau$, with potential
\$ U_\tau = (ev)^2. \$
The constraints on~$\Bmu$ allow two polarizations:
\be
\eps_{\tau1} &=& (p,\w,0,0)/E \label{define-tau1} \\
\eps_{\tau2} &=& (0,0,1,0).   \label{define-tau2}
\ee
We name the two corresponding scalar modes $B_{\tau1}$ and~$B_{\tau2}$.

To obtain a third scalar mode from equation~\(eom-B)
we make the reasonable assumption that
the polarization of the third mode
is orthogonal to the polarizations of the two we already have.
Referring back to equations \(define-tau1)~and~\(define-tau2)
we find that orthogonality requires
the first three components of the polarization of the third mode
to be proportional to~$(\w,p,0)$,
so we can write
\$ \Bmu = \d^\mu b \qquad\hbox{for~$\mu \ne 3$.} \label{define-b} \$
Now the four components of equation~\(eom-B) reduce to
the following two equations:
\be
[-\d_3^2 + (ev)^2]\rlap{\hskip0.7ex $b$}\hphantom{B^3} &=& \d_3 B^3 \\ \relax
[-E^2    + (ev)^2]B^3 &=& E^2 \d_3 b.
\ee
Solving for~$b$ in terms of~$B^3$ yields\footnote{%
  We must eliminate $b$ rather than~$B^3$, for the following heuristic reason:
  Suppose we send a particle with the third polarization toward the wall
  with energy low enough that it will be totally reflected.
  When the particle reaches the classical turning point
  its momentum is~$(\w,p,0,0)$.
  Now, for a massive vector boson with this momentum
  the third polarization is~$(0,0,0,1)$,
  so at this particular point in the scattering
  $b$ will be forced to zero,
  not by the differential equation it obeys,
  but by the changing polarization vector.
  $B^3$ has no such singularity.}
\$ b = {1\over E^2 (ev)^2} \d_3 [ (ev)^2 B^3 ]; \label{solve-for-b} \$
the remaining equation of motion is
\$ \biggl[ - E^2 - \d_3^2 - 2\biggl(\vpv\biggr)\d_3
           + (ev)^2 + 2\biggl(\vpv\biggr)^2 - 2\vppv \biggr] B^3 = 0. \$
Defining
\$ B^3 = (E/ev)\l \label{define-lambda} \$
removes the linear derivative term, leaving
\$ \biggl[ - E^2 - \d_3^2
           + (ev)^2 + 2\biggl(\vpv\biggr)^2 - \vppv \biggr] \l = 0,
   \label{eom-l} \$
a scalar scattering equation with potential
\$ U_\l = (ev)^2 + 2\biggl(\vpv\biggr)^2 - \vppv. \$
This scalar mode, which describes the third polarization of the field~$\Bmu$,
we refer to by the name~$B_\l$.

\isubsection{Connection of internal and asymptotic modes}
\label{s:connection}

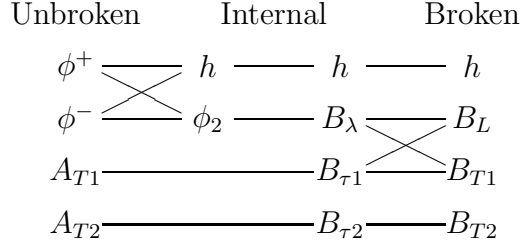
\begin{figure}\centered
\begin{picture}(170,100)
\thinlines
\multiput(20,10)(0,20){2}{\line(1,0){80}}
\multiput(20,50)(0,20){2}{\line(1,0){30}}
\multiput(70,50)(0,20){2}{\line(1,0){30}}
\multiput(120,10)(0,20){4}{\line(1,0){30}}
\put(20,52.5){\line(2,1){30}}
\put(20,67.5){\line(2,-1){30}}
\put(120,32.5){\line(2,1){30}}
\put(120,47.5){\line(2,-1){30}}
\put(10,10){\makebox(0,0){$A_{T2}$}}
\put(10,30){\makebox(0,0){$A_{T1}$}}
\put(10,50){\makebox(0,0){$\pmi$}}
\put(10,70){\makebox(0,0){$\ppl$}}
\put(60,50){\makebox(0,0){$\p2$}}
\put(60,70){\makebox(0,0){$h$}}
\put(110,10){\makebox(0,0){$B_{\tau2}$}}
\put(110,30){\makebox(0,0){$B_{\tau1}$}}
\put(110,50){\makebox(0,0){$B_\l$}}
\put(110,70){\makebox(0,0){$h$}}
\put(160,10){\makebox(0,0){$B_{T2}$}}
\put(160,30){\makebox(0,0){$B_{T1}$}}
\put(160,50){\makebox(0,0){$B_L$}}
\put(160,70){\makebox(0,0){$h$}}
\put(10,90){\makebox(0,0){Unbroken}}
\put(85,90){\makebox(0,0){Internal}}
\put(160,90){\makebox(0,0){Broken}}
\end{picture}
\caption{Connections between internal and asymptotic modes.}
\label{f:connection}
\end{figure}

A schematic diagram of how the various modes connect to one another
is given in figure~\ref{f:connection}.
We will devote the rest of this section
to making a precise statement of what this diagram means.

The fields~$h$ and~$\Bmu$ which describe the scattering process
are the same as the fields in the broken phase,
so to find the connection between modes at~$\pinf$
we need only express the asymptotic polarizations of~$\Bmu$
in terms of the internal ones.
We summarize the various polarization vectors in table~\ref{t:polarizations}.
The only new information in this table
is the polarization vector~$\eps_\l$.
To calculate this vector,
we consider a scalar function~$\l$ which is asymptotically a plane wave:
\$ \l = \eipx. \$
Working backward through equations \(define-lambda),
\(solve-for-b), and~\(define-b) we find
\be
B^3   &=& \biggl({E\over ev}\biggr) \eipx \label{lambda-B3} \\
b     &=& {1 \over E(ev)} \biggl(ik + \vpv\biggr) \eipx \\
B^\mu &=& {p^\mu \over E(ev)} \biggl(k-i\vpv\biggr) \eipx. \label{lambda-Bmu}
\ee
Using the fact that $v'/v \to 0$ in the broken phase,
we read off the unnormalized polarization vector
from equations \(lambda-B3) and~\(lambda-Bmu).
\begin{table}
$$ \begin{array}{|c|cccc|c|} \cline{1-6}
p^\mu&\w&p&0&k&\\ \cline{1-6}
\eps_L&\openabove{{\w^2-M^2\over\w}}&p&0&k&{\w\over M\sqrt{\w^2-M^2}}\\
\eps_{T1}&0&k&0&-p&\openabove{{1\over \sqrt{p^2+k^2}}}\\
\eps_{T2}&0&0&1&0&\\ \cline{1-6}
\eps_\l&\w&p&0&\openabove{{E^2\over k}}&{k\over ME}\\
\eps_{\tau1}&p&\w&0&0&\openabove{\f1/E}\\
\eps_{\tau2}&0&0&1&0&\vphantom{\openabove{\f1/E}}\\ \cline{1-6}
\end{array} $$
\caption{Summary of polarization vectors.
The vector~$\eps_\l$ is only meaningful in the broken phase.
The last column contains normalization factors
for the normalization~$\eps^2 = -1$.}
\label{t:polarizations}
\end{table}

Expanding the broken-phase polarization vectors in terms of the internal ones
we obtain the following matrix equation
connecting the internal modes to the broken-phase modes:
\$
\def\vph{\vphantom{\pM}}
\cvec{h\\B_L\vph\\B_{T1}\vph\\B_{T2}} =
\matr{1&\0&\0&\0\\
      \0&\ghost\wk&\pM&\0\\
      \0&-\pM&\wk&\0\\
      \0&\0&\0&1}
\cvec{h\\B_\l\vph\\B_{\tau1}\vph\\B_{\tau2}}, \label{connect-broken}
\$
where zeroes in the matrix have been replaced by blanks for legibility.
At normal incidence the matrix reduces to the identity.

The connection between the internal modes
and the asymptotic modes in the unbroken phase is more involved.
We begin with the simplest case,
the connection between $B_{\tau2}$ and~$A_{T2}$.
We consider a solution of the scattering equation for the $B_{\tau2}$~mode
which is asymptotically a plane wave,
that is, for which there is a complex number~$B_{\tau2}$ such that
\$ \tau - B_{\tau2}\eipx \to 0 \_as_ z\to\minf. \$
According to the definition~\(define-tau) of~$\tau$,
this solution generates a field configuration
\$ \Bmu[B_{\tau2}] = \eps_{\tau2}^\mu \tau. \$
On the other hand,
from the definition~\(define-B) of~$\Bmu$ we find that
the field configuration generated by the standard solution
for the $A_{T2}$~mode is
\$ \Bmu[A_{T2}] = A_{T2}\,\eps_{T2}^\mu \eipx. \$
If we take $A_{T2}$ equal to~$B_{\tau2}$,
the two field configurations are asymptotically the same, that is,
\$ \Bmu[B_{\tau2}] - \Bmu[A_{T2}] \to 0 \_as_ z\to\minf. \$
We use this condition%
---that the difference between field configurations go to zero---%
as our criterion for connection.

We handle the $B_{\tau1}$~mode similarly.
An internal solution with asymptotic form
\$ \tau = B_{\tau1}\eipx \$
generates an asymptotic field configuration
\$ \Bmu[B_{\tau1}] = B_{\tau1}\,\eps_{\tau1}^\mu \eipx. \$
Applying the relation
\$ \eps_{\tau1}^\mu = (\sgn k)\,\eps_{T1}^\mu
    + \biggl({p \over \w E}\biggr) p^\mu \$
separates $\Bmu[B_{\tau1}]$ into two terms.
The first term matches the field~$\Bmu[A_{T1}]$
generated by an $A_{T1}$~solution with amplitude~$A_{T1}=(\sgn k)B_{\tau1}$.
The second term describes a field
\$ \Amu = B_{\tau1}\,\biggl({p \over \w E}\biggr) p^\mu \eipx
   \label{pure-gauge} \$
polarized in the direction of~$p^\mu$.
Although the field~\(pure-gauge) is a valid solution
of the unbroken-phase equation of motion~\(unbroken-eom-A) for~$\Amu$,
it has a field strength~($F^{\mu\nu}$) of zero
and so carries no energy or momentum.
We discard this pure gauge solution
and find that $B_{\tau1}$~connects directly to~$A_{T1}$.

The use of the coefficients $B_{\tau1}$~and~$B_{\tau2}$
in the preceding paragraphs
has obeyed the following convention, which we will continue to observe:
whenever we use a complex number to describe
the asymptotic form of one of the internal modes,
the number is the coefficient of a unit-amplitude plane wave.
We will apply the same convention to $h$~and~$\p2$
even though they are not internal modes.

Next we discuss the connection between the internal mode~$B_\l$
and the asymptotic modes in the unbroken phase.
A solution of the scattering equation
which is asymptotically a plane wave,
that is, for which
\$ \l - B_\l \eipx \to 0 \_as_ z\to\minf, \$
generates an asymptotic field configuration
\$ B^3[B_\l] = B_\l \biggl({E\over ev}\biggr) \eipx. \$
Since $v\to 0$ in the unbroken phase,
this field configuration is divergent.
However, we can obtain an equal divergence from the asymptotic field~$\p2$.
We take $\p2$ to be a plane wave
with amplitude given by the complex number~$\p2$
and substitute into the definition~\(define-B) of~$\Bmu$ to find
\$ B^3[\p2] = \p2 \biggl({1\over ev}\biggr) (\vpv-ik) \eipx. \$
As~$v \to 0$, the ratio~$v'/v$ reduces to~$m$,
so we can match the two divergences by taking
\$ \p2 = {E\over  m-ik} B_\l = {m+ik\over E} B_\l. \label{connect-p2-l} \$

Although equation~\(connect-p2-l) will turn out to be
the correct relation between the modes $B_\l$~and~$\p2$,
the above argument using matching divergences
does not prove that the two modes connect,
because the difference of the matching divergences contains a finite part
which does not necessarily go to zero as~$z\to\minf$.
To calculate the finite part of the field~$\Bmu[B_\l]$ correctly,
we keep the first-order term
in the asymptotic expansion of~$\l$ near~$v=0$, writing
\$ \l = B_\l (1 + cv) \eipx, \$
where the constant~$c$ is determined by
the scattering equation~\(eom-l) for the $B_\l$~mode.
A calculation which we omit shows that
when the relation between the modes $B_\l$~and~$\p2$
is that given in equation~\(connect-p2-l),
the field configurations $\Bmu[B_\l]$~and~$\Bmu[\p2]$ satisfy
\$ \Bmu[B_\l] - \Bigl( \Bmu[\p2] + C\,p^\mu \eipx \Bigr) \to 0
   \_as_ z\to 0, \$
where the coefficient~$C$ of the pure gauge solution
is an appropriately chosen constant.
In other words, the field configuration~$\Bmu[B_\l]$
is asymptotically equal to the field~$\Bmu[\p2]$ plus a pure gauge term.
We discard the pure gauge solution and find that $B_\l$ connects to~$\p2$
as in equation~\(connect-p2-l).

All that remains is to combine the $\p2$ produced by~$B_\l$
with the internal mode~$h$
to get the unbroken-phase modes $\ppl$ and~$\pmi$.
Since $h$ and~$\p2$ are real-valued fields,
the amplitudes $h$ and~$\p2$ actually represent the functions
\$ {h  \over 2}\eipx + {h^*  \over 2}\epipx \$
and
\$ {\p2\over 2}\eipx + {\p2^*\over 2}\epipx \$
which we combine into
\$ \phi[h,\p2] = \biggl({h  +i\p2  \over\rt}\biggr) (\f1/2 \eipx)
               + \biggl({h^*+i\p2^*\over\rt}\biggr) (\f1/2 \epipx). \$
Using the normalized solutions~\(unbroken-solutions)
we read off the amplitudes
\$ \ppl = {h+i\p2\over\rt} \_and_ \pmi = {h-i\p2\over\rt}.
   \label{particle-antiparticle} \$

We collect all our results
about the connection between the internal and unbroken-phase modes
in the following equation:
\$
\def\factor{{k-im\over E\rt}}
\def\vph{\vphantom{\factor}}
\cvec{\ppl\vph\\\pmi\vph\\A_{T1}\\A_{T2}} =
\matr{1/\rt&-\factor&\0&\0\\
      1/\rt&\ghost\factor&\0&\0\\
      \0&\0&\sgn k&\0\\
      \0&\0&\0&1}
\cvec{h\vph\\B_\l\vph\\B_{\tau1}\\B_{\tau2}}. \label{connect-unbroken}
\$
The matrix equations \(connect-broken)~and~\(connect-unbroken)
are the desired precise statement
of what the connection diagram in figure~\ref{f:connection} means.

\section{Scattering from the domain wall}
\label{sec:scatter}
\isubsection{Specification of the potential}

All our results up to this point are independent
of the details of the Higgs potential~$V(v)$.
To obtain anything more than generalities
about scattering from the domain wall, however,
it is necessary to give up this independence
and specify a form for the potential.
We choose the standard quartic potential
\$ V(v) = \f1/4 \l v^2(v-\vo)^2. \label{quartic} \$
In section~\ref{s:discussion}
we will discuss how our results depend on this choice.

With this choice of potential
the mass~$m$~\(mass-charged-scalars) of the charged scalars
is equal to the mass~$\mtilde$~\(mass-higgs) of the Higgs,
\$ m^2 = \mtilde^2 = \f1/2 \l \vo^2, \$
and the integral~\(vev-integral) which determines the background field
can be evaluated explicitly, with result
\$ v(z) = \f\vo/2 \biggl( 1+\tanh {mz\over 2} \biggr). \$
The scattering potentials $U_h$, $U_\tau$ and~$U_\l$
can be cast into the standard form
\$ U = m^2 \Bigl[ U_0 (1-s) + U_1 s + U_2 s(1-s) \Bigr],
   \label{standard-form} \$
where the dimensionless coordinate~$s$ is defined by
\$ s = \f1/2 \biggl( 1+\tanh {mz\over 2} \biggr) = {v\over\vo}.
   \label{s-tanh} \$
The coefficients $U_0$, $U_1$ and~$U_2$ for the various potentials
are listed in table~\ref{t:potentials}.
\begin{table}
$$ \begin{array}{|c|rcl|c|ccc|} \cline{2-8}
\mc&\multicolumn{3}{|c|}{\hbox{General~$V$}}&\hbox{Specific~$V$}&
  \mskip7mu U_0\mskip7mu&\mskip4mu U_1\mskip4mu&\mskip1mu U_2\\ \cline{1-8}
h&U_h&=&\openabove{V''(v)}&\multicolumn{1}{l|}{m^2[\mskip1mu
  1-6s(1-s)\mskip1mu]}&1&1&-6\\
B_\tau&U_\tau&=&(ev)^2&\multicolumn{1}{r|}{M^2s^2}&0&\M^2&-\M^2\\
B_\l&U_\l&=&\openbelow{(ev)^2+2(\vpv)^2-\vppv}&m^2(1-s)+M^2s^2&
  1&\M^2&-\M^2\\ \cline{1-8}
\end{array} $$
\caption{Summary of internal mode scattering potentials.
The mass ratio~$\M^2$ is defined to be~$M^2/m^2$.}
\label{t:potentials}
\end{table}

\isubsection{Solution using hypergeometric functions}
\label{s:solution}

In this section we solve the scattering equation
$$ [\dbox + U(z)]\,S = 0 \eqno{\(eom-scalar)} $$
analytically for any potential~$U$ of the standard form~\(standard-form),
that is, any potential quadratic in the variable~$s$.
Although our results are more general,
our method of solution is similar to that used by Ayala et al~\cite{ayala}
to study the scattering of fermions.
We factor~$S$ into
\$ S = s^\k0 (1-s)^\k1 \, \tilde S(s) \> \eperp, \$
where $\k0$ and~$\k1$ are constants to be determined,
and find that the scattering equation becomes
\ble
\lefteqn{ \biggl[ \; s^2 (1-s)^2 {d^2\over ds^2} } \no
&+& \biggl( 2\k0 \, s (1-s)^2 - 2\k1 \, s^2 (1-s)
   + s(1-s)(1-2s) \biggr) {d\over ds} \no
&+& \biggl( \k0^2 \, (1-s)^2 + \k1^2 \, s^2
   -  (\k0 + \k1 + 2\k0\k1) \, s(1-s) \no
 && \hphantom{\biggl( \k0^2 \, (1-s)^2 + \k1^2 \,
   s^2 - (\k0 + \k1 + 2\k0\k1) \, s(1-s)}
    \llap{$\displaystyle {} + {E^2\over m^2} - U_0 \,
(1-s) - U_1 \, s - U_2 \, s(1-s)$}
    \; \biggr) \; \biggr] \; \tilde S = 0. \quad \label{hg-almost}
\ele
Setting
\$ \k{i}^2 = U_i - {E^2\over m^2} \label{kappa-def} \$
makes equation~\(hg-almost) divisible by~$s(1-s)$;
the quotient is the hypergeometric equation\footnote{%
  All the relevant facts about the hypergeometric equation
  appear in reference~\cite{landau}.}
\$
  \biggl[ s(1-s) {d^2\over ds^2}
+ \biggl( \gamma - (\alpha + \beta + 1) s \biggr) {d\over ds}
- \alpha\beta \biggr] \tilde S = 0 \label{hg}
\$
with parameters
\be
\alpha & = & \k0 + \k1 + \textf1/2 + \delta \no
\beta  & = & \k0 + \k1 + \textf1/2 - \delta \label{hg-params} \\
\gamma & = & 1 + 2\k0. \nonumber
\ee
The constant~$\delta$ is defined by
\$ \delta = \sqrt{1/4 - U_2}. \$

One solution of the hypergeometric equation~\(hg) is
the hypergeometric function~$F(\alpha,\beta,\gamma;s)$,
and the corresponding solution of the scattering equation~\(eom-scalar) is
\$ S = s^\k0 (1-s)^\k1 \, F(\alpha,\beta,\gamma;s) \> \eperp.
   \label{hg-solution} \$
We study the asymptotic behavior of this solution,
beginning with the case $z\to\minf$, that is, $s\to 0$.
As its argument~$s$ goes to zero the function~$F$ reduces to unity,
as does the factor~$(1-s)^\k1$.
{}From equation~\(s-tanh) we find that as~$z\to\minf$
the coordinate~$s$ simplifies to~$\exp{mz}$,
so the solution~\(hg-solution) is asymptotically the plane wave
\$ \exp{-i\w t + ipx + m\k0 z}. \label{hg-asy-minf} \$
For the solution to describe scattering,
this exponential must represent
the transmitted part of a wave incident from~$z=\pinf$,
that is, $\k0$~must be negative imaginary.
To avoid confusion, we take $\k1$~to be negative imaginary as well.

To calculate the asymptotic behavior
of the solution~\(hg-solution) in the opposite limit~$z\to\pinf$
we use the identity
\ble
\lefteqn{ \displaystyle
F(\alpha,\beta,\gamma;s) =
  {\Gamma(\gamma)\Gamma(\gamma-\alpha-\beta)
\over\Gamma(\gamma-\alpha)\Gamma(\gamma-\beta)}
  F(\alpha,\beta,\alpha+\beta+1-\gamma;1-s) } \\
\noalign{\vskip1.5ex}
\displaystyle
&+& {\Gamma(\gamma)\Gamma(\alpha+\beta-\gamma)
\over\Gamma(\alpha)\Gamma(\beta)}
  (1-s)^{\gamma-\alpha-\beta}
F(\gamma-\alpha,\gamma-\beta,\gamma+1-\alpha-\beta;1-s). \nonumber
\ele
Taking into account the parameter values~\(hg-params)
and replacing $1-s$~by its equivalent~$\exp{-mz}$
we find $S$ has the following asymptotic form at~$\pinf$:
\ble
 && \overalign{\Gamma(1+2\k0)}{\,\Gamma(-2\k1)}%
              {\Gamma(\k0-\k1+\f1/2+\delta)}{\,
\Gamma(\k0-\k1+\f1/2-\delta)} \, \exp{-i\w t + ipx - m\k1 z} \no
&+& \overalign{\Gamma(1+2\k0)}{\,\Gamma( 2\k1)}%
              {\Gamma(\k0+\k1+\f1/2+\delta)}{\,
\Gamma(\k0+\k1+\f1/2-\delta)} \, \exp{-i\w t + ipx + m\k1 z}.
\label{hg-asy-pinf}
\ele
Since $\k1$~is negative imaginary,
the second term represents the incident wave.
Taking the appropriate ratios yields
the reflection and transmission coefficients
\be
r &=& \overalign{\Gamma(\k0+\k1+\f1/2+\delta)}{\,
\Gamma(\k0+\k1+\f1/2-\delta)}%
                {\Gamma(\k0-\k1+\f1/2+\delta)}{\,
\Gamma(\k0-\k1+\f1/2-\delta)} \;
\overalign{}{\Gamma(-2\k1)}{}{\Gamma(\ghost 2\k1)} \label{refl-ampl} \\
t &=& \overalign{\Gamma(\k0+\k1+\f1/2+\delta)}{\,
\Gamma(\k0+\k1+\f1/2-\delta)}%
{\Gamma(1+2\k0)}{\,\Gamma(2\k1)}. \label{trans-ampl}
\ee

In the case of total reflection
we make the transmitted wave~\(hg-asy-minf)
into a decaying exponential by taking $\k0$~to be positive real.
Although the particle is totally reflected,
the transmission coefficient~$t$ does not become zero,
and indeed it need not,
being merely the coefficient of a decaying exponential.
The reflection coefficient~$r$ does reach unit amplitude, however.

The solution of the hypergeometric equation~\(hg)
which represents a particle incident from~$\minf$ is
\$ \tilde S = F(\alpha,\beta,\alpha+\beta+1-\gamma;1-s). \$
Proceeding as before
we find reflection and transmission coefficients
which differ from the previous results \(refl-ampl)~and~\(trans-ampl)
only by exchange of $\k0$~and~$\k1$.

\isubsection{Bound states}
\label{s:bound-states}

To search for bound states
we take both~$\k0$ and~$\k1$ to be positive real
and require that the growing exponential in equation~\(hg-asy-pinf)
have coefficient zero:
\$ \overalign{\Gamma(1+2\k0)}{\,\Gamma( 2\k1)}%
             {\Gamma(\k0+\k1+\f1/2+\delta)}{\,
\Gamma(\k0+\k1+\f1/2-\delta)} \, = 0. \label{no-grow} \$
The gamma function has no zeroes,
but it does have a pole at each nonpositive integer.
The parameter~$\alpha = \k0+\k1+\f1/2+\delta$
is always either strictly positive or complex,
so equation~\(no-grow) reduces to the condition that
the parameter~$\beta = \k0+\k1+\f1/2-\delta$ be a nonpositive integer,
that is, that
\$ -\beta \equiv \delta-\textf1/2-(\k0+\k1) = n \label{minus-beta} \$
for some nonnegative integer~$n$.
We attempt to satify this condition by varying the energy~$E^2$.
Our variations must keep $\k0$~and~$\k1$ real-valued and strictly positive,
which, according to the definition~\(kappa-def),
means that
\$ {E^2\over m^2} < \min\,(U_0,U_1) \$
and therefore that
\$ \k0 + \k1 > \sqrt{|U_0-U_1|}. \$
As a result, $-\beta$~may take any value from~$\minf$
up to but not including a maximum~$N$ given by
\$ N = \delta - \textf1/2 - \sqrt{|U_0-U_1|}, \label{bound-N} \$
and equation~\(minus-beta) can be satisfied
for any nonnegative integer~$n < N$.
Solving equation~\(minus-beta) for~$E^2$
yields the energy of the corresponding bound state
\$ {E^2\over m^2} = \min\,(U_0,U_1)
                  - \left( {(\delta - \f1/2 - n)^2
- |U_0-U_1|\over 2(\delta - \f1/2 - n)} \right)^2.
\label{bound-energy} \$
For bound states the solution~\(hg-solution)
of the hypergeometric equation simplifies,
because when $-\beta$~is equal to a nonnegative integer~$n$
the power series for the hypergeometric function
\$ F(\alpha,\beta,\gamma;s) = 1 + {\alpha\beta\over\gamma} s
   + {\alpha(\alpha+1)\beta(\beta+1)\over\gamma(\gamma+1)}
     {s^2\over 2!} + \cdots \$
collapses to a polynomial of degree~$n$.
The lowest-energy bound state has the particularly simple wavefunction
\$ S = s^\k0 (1-s)^\k1 \>\eperp. \label{ground-state} \$

Using the parameters~$U_i$ given in table~\ref{t:potentials}
we can compute the number of bound states for each of the internal modes.
For the $B_\tau$~mode we find the bound state parameter~$N$ is never positive,
so there are no bound states,
but then we expect none in a potential without a minimum.

The $h$~mode has $N = 2$ and so has two bound states.
We apply equations~\(bound-energy)--\(ground-state) to find
that the first bound state has energy~$E^2 = 0$ and wavefunction
\$ h = s(1-s) \>\eperp = \f1/4 \sech^2 {mz\over 2} \>\eperp
   \label{bound-h-first} \$
and that the second has energy~$E^2 = (3/4) m^2$ and wavefunction
\$ h = \sqrt{s(1-s)}\;(1-2s) \>\eperp
     = -\f1/2 \sech {mz\over 2} \tanh {mz\over 2} \>\eperp.
   \label{bound-h-second} \$
To make a physical interpretation of these bound states
we remember that $h$~represents a small perturbation
added to the background field~$v$.
The factor~$\sech^2(mz/2)$ which appears
in the first bound state~\(bound-h-first) is proportional to~$v'(z)$,
and adding a small amount of~$v'(z)$ to~$v$ has the effect of
translating the wall in the $z$-direction,
so the wavefunction~\(bound-h-first) describes
a plane-wave oscillation of the local position of the wall.
The second bound state~\(bound-h-second) describes
a similar oscillation of the local thickness of the wall.
These two excitations of the wall can be thought of
as quasiparticles confined to the wall surface
with masses~$\w^2 - p^2 = E^2$ of zero and~$(3/4)m^2$ respectively.
No matter what Higgs potential~$V$ we choose,
the $h$~mode always has a bound state
for oscillations of the wall position---%
we need only apply the equation of motion~\(eom-v) for~$v$
to find that the wavefunction~$h = v'(z)\eperp$ satisfies equation~\(eom-h)
whenever~$E^2 = 0$.
The underlying physical reason for the existence of this bound state
is the translation invariance of the Lagrangian.

\begin{figure}\centered
\epsf{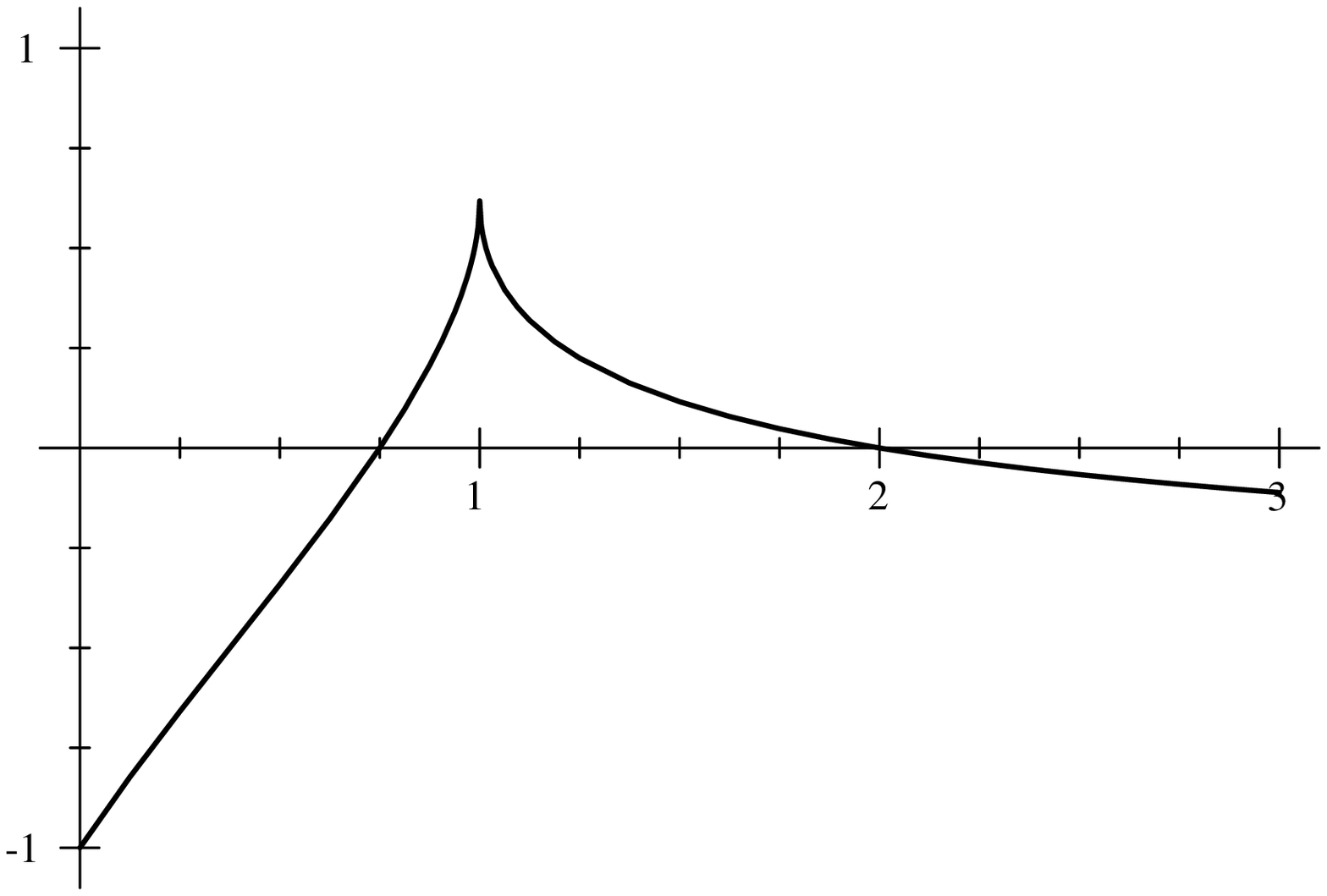}{2.5in}
\begin{picture}(0,0)
\put(9,91){\makebox(0,0){$\M\rlap{$\smash{^2}$}$}}
\put(-256,191){\makebox(0,0){$N$}}
\end{picture}
\epsf{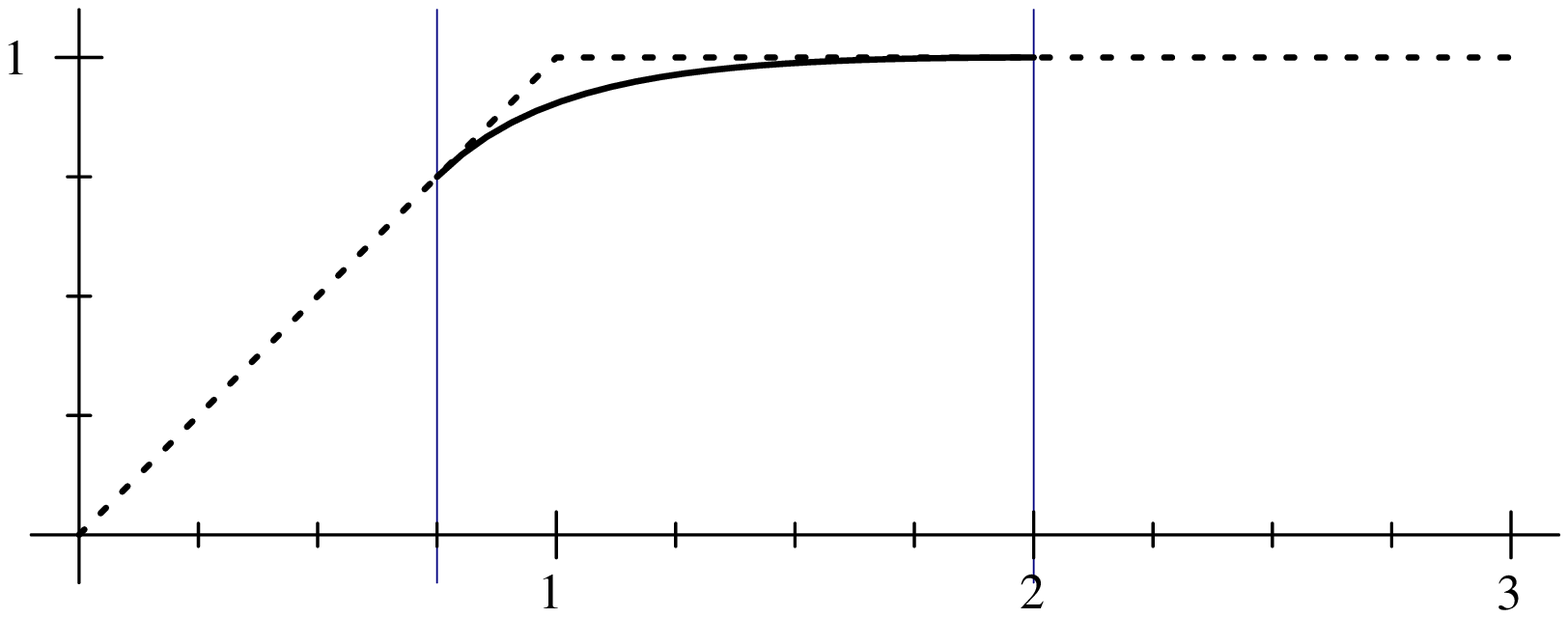}{2.5in}
\begin{picture}(0,0)
\put(9,51){\makebox(0,0){$\M\rlap{$\smash{^2}$}$}}
\put(-256,152){\makebox(0,0){$E^2/m^2$}}
\end{picture}
\vskip-2.8em
\caption{Bound state parameter~$N$ and bound state energy~$E^2/m^2$
as functions of~$\M^2$ for the $B_\l$~mode.
The dotted line in the second plot marks the start of
the continuum of unbound states.}
\label{f:boundstate}
\end{figure}

For mass ratios~$\M^2 = M^2/m^2$ in the range~$3/4 < \M^2 < 2$
the $B_\l$~mode has a single bound state.
Neither the energy~\(bound-energy) nor the wavefunction~\(ground-state)
of this state simplify further.
The bound state parameter~$N$ and energy~$E^2$ are plotted
as functions of~$\M^2$ in figure~\ref{f:boundstate}.

\isubsection{Reflection and transmission probabilities}
\label{s:scatter-internal}

For the case in which total reflection does not occur
we calculate the reflection and transmission probabilities $R$~and~$T$
for a particle incident from~$\pinf$.
Starting from the coefficients~\(refl-ampl) and~\(trans-ampl)
and applying the identity
\$ \Gamma(z)\Gamma(1-z) = {\pi\over\sin\pi z} \$
we find
\be
R = |r|^2 &=&
\overalign{\sin\pi(\k0-\k1+\f1/2+\delta)}{\,\sin\pi(\k0-\k1+\f1/2-\delta)}%
          {\sin\pi(\k0+\k1+\f1/2+\delta)}{\,\sin\pi(\k0+\k1+\f1/2-\delta)}
\label{refl-prob} \\
T = {k_0\over k_1}|t|^2 &=&
\overalign{-\sin 2\pi\k0}{\,\sin 2\pi\k1}%
          {\sin\pi(\k0+\k1+\f1/2+\delta)}{\,\sin\pi(\k0+\k1+\f1/2-\delta)}
\label{trans-prob}
\ee
where the asymptotic momenta~$k_i$ are given by
\$ k_i = \sqrt{E^2 - m^2U_i} = im\k{i}. \label{define-k} \$
The probabilities \(refl-prob)~and~\(trans-prob)
are invariant under exchange of $\k0$~and~$\k1$
and so apply to particles incident from~$\minf$ as well.

\begin{figure}\centered
\epsf{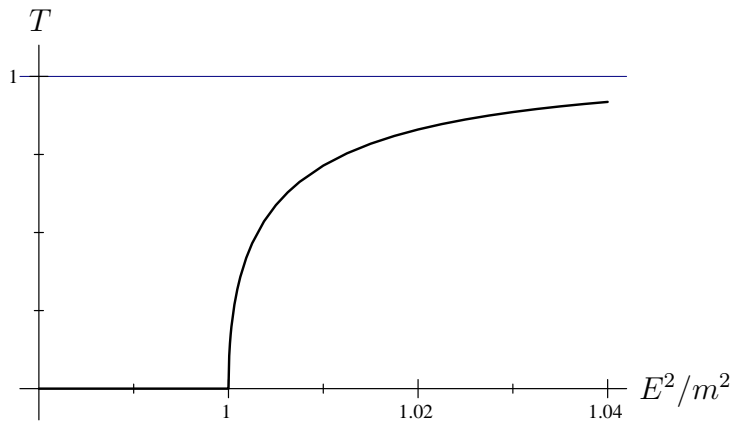}{2in}
\begin{picture}(0,0)
\put(22,13){\makebox(0,0){$E^2/m^2$}}
\put(-222,153){\makebox(0,0){$T$}}
\end{picture}
\caption{Transmission probability for the $B_\tau$~mode when~$\M^2 = 1$.}
\label{f:transmission}
\end{figure}

Using trigonometric identities we convert the transmission probability
from equation~\(trans-prob) into the more practical form
\$
\def\x#1{{2\pi k_{#1}\over m}}
T = \overalign{2\,}{\sinh\x0 \sinh\x1}{\cosh\x0 \cosh\x1
+ {}}{\sinh\x0 \sinh\x1 + \cos 2\pi \delta}.
\$
Except in a few cases which we discuss momentarily,
the transmission probability is
a strictly increasing function of the energy~$E^2$
with a square-root singularity at the threshold of total reflection.
A typical curve is shown in figure~\ref{f:transmission}.
When $U_0 = U_1$ the square-root singularity
is replaced by a linear approach to zero,
unless also $\cos 2\pi \delta = -1$,
in which case the transmission probability~$T \equiv 1$
is no longer even strictly increasing.
According to figure~\ref{f:transmission}
the energy~$E^2$ need only be a
small fraction of the mass~$m^2$ above threshold
for the transmission probability to be significant.
Other choices of the scattering potential
do not make the onset of transmission significantly slower---%
in all cases the transmission probability
reaches~$1/2$ before~$E^2$ gets~$0.02 \, m^2$ above the threshold.
On the other hand, the onset becomes much faster
as the potential approaches any of the potentials which give~$T \equiv 1$.

The transmission curves for the $B_\tau$ and~$B_\l$~modes
have the generic form described in the previous paragraph.
If the gauge boson mass~$M$ were zero,
the $B_\tau$~mode would have $T \equiv 1$,
so if the gauge boson mass~$M$ is small compared to~$m$
the curve for the $B_\tau$~mode will have a sharp onset.
The onset in the $B_\l$~mode is never particularly sharp.
For the $h$~mode, $U_0 = U_1$ and $\delta = 5/2$
make the transmission probability {\em unity}.
This result does not generalize to an arbitrary Higgs potential~$V$.
For most choices of~$V$,
the mass~$m$ of the charged scalars
is different from the mass~$\mtilde$ of the Higgs,
so that reflection---in fact, total reflection---is possible.

\isubsection{Scattering of asymptotic modes}
\label{s:scatter-asymptotic}

In section~\ref{s:connection}
we derived the connection equations
$$
\def\vph{\vphantom{\pM}}
\cvec{h\\B_L\vph\\B_{T1}\vph\\B_{T2}} =
\matr{1&\0&\0&\0\\
      \0&\ghost\wk&\pM&\0\\
      \0&-\pM&\wk&\0\\
      \0&\0&\0&1}
\cvec{h\\B_\l\vph\\B_{\tau1}\vph\\B_{\tau2}} \eqno{\(connect-broken)}
$$
and
$$
\def\factor{{k-im\over E\rt}}
\def\vph{\vphantom{\factor}}
\cvec{\ppl\vph\\\pmi\vph\\A_{T1}\\A_{T2}} =
\matr{1/\rt&-\factor&\0&\0\\
      1/\rt&\ghost\factor&\0&\0\\
      \0&\0&\sgn k&\0\\
      \0&\0&\0&1}
\cvec{h\vph\\B_\l\vph\\B_{\tau1}\\B_{\tau2}}. \eqno{\(connect-unbroken)}
$$
Using these two equations
and the reflection and transmission coefficients
\(refl-ampl)~and~\(trans-ampl)
we can calculate scattering probabilities
for a particle incident from the broken or unbroken phase.
We use an inverted connection equation
to convert a unit amplitude for the incident asymptotic mode
into amplitudes for internal modes,
multiply by the appropriate reflection and transmission coefficients
to find the scattered amplitudes at both positive and negative infinity,
and use the connection equations
to convert back to amplitudes for asymptotic modes.
We take the norm of these amplitudes to get probabilities,
multiplying in the case of transmission by the appropriate momentum ratio.

The meaning of the symbol~$k$
depends on the context in which it appears.
To avoid confusion,
we define for each of the different particle masses
a corresponding positive momentum,
for example,
\$ k_m = \sqrt{E^2 - m^2}, \$
and write each~$k$ in terms of these momenta.
In the connection equation~\(connect-broken) for the broken phase,
$k$~is the momentum of a massive vector boson, that is, $|k| = k_M$.
Because equation~\(connect-broken) expresses asymptotic amplitudes
in the broken phase
in terms of internal ones,
we will apply it to particles moving away from the wall,
that is, particles with positive $z$-momentum.
Setting $k = +k_M$ we obtain
\$
\def\vph{\vphantom{\xpM}}
\cvec{h\\B_L\vph\\B_{T1}\vph\\B_{T2}} =
\matr{1&\0&\0&\0\\
      \0&\ghost\xwk&\xpM&\0\\
      \0&-\xpM&\xwk&\0\\
      \0&\0&\0&1}
\cvec{h\\B_\l\vph\\B_{\tau1}\vph\\B_{\tau2}}. \label{connect-broken-k}
\$
To obtain the connection equation for particles moving toward the wall,
we invert equation~\(connect-broken) and take $k = -k_M$, with result
\$
\def\vph{\vphantom{\xpM}}
\cvec{h\\B_\l\vph\\B_{\tau1}\vph\\B_{\tau2}} =
\matr{1&\0&\0&\0\\
      \0&-\xwk&-\xpM&\0\\
      \0&\ghost\xpM&-\xwk&\0\\
      \0&\0&\0&1}
\cvec{h\\B_L\vph\\B_{T1}\vph\\B_{T2}}. \label{inverted-broken-k}
\$
Substituting $k = -k_m$ into equation~\(connect-unbroken)
gives the connection equation for particles moving away from the wall
into the unbroken phase,
\$
\def\factor{{k_m+im\over E\rt}} 
\def\vph{\vphantom{\factor}}
\cvec{\ppl\vph\\\pmi\vph\\A_{T1}\\A_{T2}} =
\matr{1/\rt&\ghost\factor&\0&\0\\
      1/\rt&-\factor&\0&\0\\
      \0&\0&-1&\0\\
      \0&\0&\0&1}
\cvec{h\vph\\B_\l\vph\\B_{\tau1}\\B_{\tau2}}, \label{connect-unbroken-k}
\$
while inverting and substituting $k = +k_m$
gives the equation for particles incident from the unbroken phase:
\$
\def\factor{{k_m+im\over E\rt}}
\def\vph{\vphantom{\factor}}
\cvec{h\\B_\l\vph\\B_{\tau1}\\B_{\tau2}} =
\matr{1/\rt&1/\rt&\0&\0\\
      -\factor&\factor&\0&\0\\
      \0&\0&1&\0\\
      \0&\0&\0&1}
\cvec{\ppl\\\pmi\vph\\A_{T1}\\A_{T2}}. \label{inverted-unbroken-k}
\$

We attach subscripts to the symbols~$r$, $t$, $R$, and~$T$
to indicate to which internal mode they apply.
As before, the symbols $r$ and~$t$
represent reflection and transmission coefficients
for a particle incident from~$\pinf$,
so for example we have
\$ T_\l = {k_m\over k_M} |t_\l|^2. \$

\begin{figure}\centered
\epsf{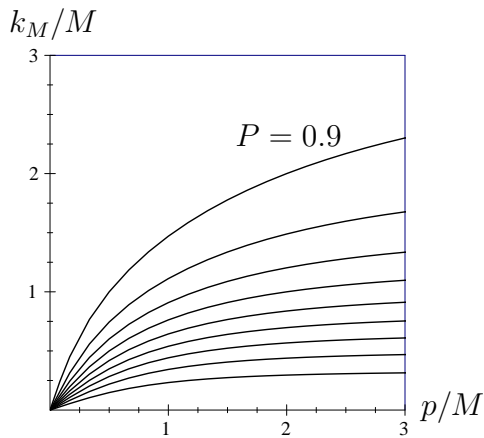}{2in}
\begin{picture}(0,0)
\put(16,10){\makebox(0,0){$p/M$}}
\put(-135,155){\makebox(0,0){$k_M/M$}}
\put(-46,113){\makebox(0,0){$P = 0.9$}}
\end{picture}
\caption{Contour plot of the normal connection probability~$P$
as a function of $p/M$ and~$k_M/M$,
with contour interval~$0.1$.
The probability~$P$ is unity at normal incidence
and zero at grazing incidence.}
\label{f:normal-connection}
\end{figure}

Using the method described at the beginning of this section,
we calculate the transmission probabilities
for a $B_L$~particle incident from the broken phase.
We follow the initial amplitude vector through
the processes of conversion into internal modes,
transmission,
and conversion into asymptotic modes at~$\minf$,
finding
\begingroup
\def\vph{\vphantom{\xwk}}
\def\factor{{k_m+im\over E\rt}}
\ble
\cvec{0\vph\\
      1\vph\\
      0\vphantom{\xpM}\\
      0}
\via(inverted-broken-k)
\cvcC{0\vph\\
      -\xwk\\
      \ghost\xpM\\
      0}
&\to&
\cvcC{0\vph\\
      -t_\l\;\xwk\\
      \ghost t_\tau\;\xpM\\
      0} \\
&\via(connect-unbroken-k)&
\cvcc{-\factor\;t_\l\;\xwk\\
      \ghost\factor\;t_\l\;\xwk\\
      \hphantom{\factor}\;{-t_\tau}\;\xpM\\
      0}. \nonumber
\ele
\endgroup
To obtain transmission probabilities
we take the norm of each element of the final vector
and multiply each by the appropriate momentum ratio.
The resulting transmission probability vector is
\$
\cvec{\f1/2\;{k_m\over k_M}|t_\l|^2\;P\\
      \f1/2\;{k_m\over k_M}|t_\l|^2\;P\\
      \hphantom{\f1/2}\;{k_0\over k_M}|t_\tau|^2\;Q\\
      0}
\equiv \cvec{\f1/2 PT_\l\vphantom{{k_m\over k_M}}\\
      \f1/2 PT_\l\vphantom{{k_m\over k_M}}\\
      \hphantom{\f1/2} QT_\tau\vphantom{{k_0\over k_M}}\\
      0}, \label{asy-trans-prob}
\$
where the probabilities $P$~and~$Q$ are
\be
P &=& {\w^2 k_M^2 \over E^2 (p^2+k_M^2)} \\
Q = 1-P &=& {p^2 M^2 \over E^2 (p^2+k_M^2)}.
\ee
The normal connection probability~$P$
ranges from unity at normal incidence to zero at grazing incidence;
its behavior at intermediate angles
is shown in figure~\ref{f:normal-connection}.

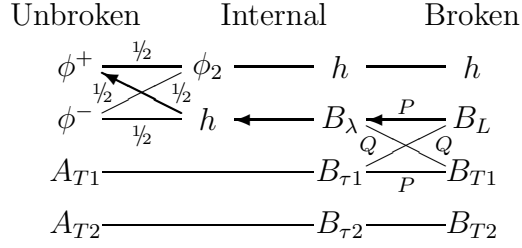
\begin{figure}\centered
\begin{picture}(170,100)
\thinlines
\multiput(20,10)(0,20){2}{\line(1,0){80}}
\multiput(20,50)(0,20){2}{\line(1,0){30}}
\multiput(70,50)(0,20){2}{\line(1,0){30}}
\multiput(120,10)(0,20){4}{\line(1,0){30}}
\put(20,52.5){\line(2,1){30}}
\put(20,67.5){\line(2,-1){30}}
\put(120,32.5){\line(2,1){30}}
\put(120,47.5){\line(2,-1){30}}
\put(10,10){\makebox(0,0){$A_{T2}$}}
\put(10,30){\makebox(0,0){$A_{T1}$}}
\put(10,50){\makebox(0,0){$\pmi$}}
\put(10,70){\makebox(0,0){$\ppl$}}
\put(60,50){\makebox(0,0){$h$}}
\put(60,70){\makebox(0,0){$\p2$}}
\put(110,10){\makebox(0,0){$B_{\tau2}$}}
\put(110,30){\makebox(0,0){$B_{\tau1}$}}
\put(110,50){\makebox(0,0){$B_\l$}}
\put(110,70){\makebox(0,0){$h$}}
\put(160,10){\makebox(0,0){$B_{T2}$}}
\put(160,30){\makebox(0,0){$B_{T1}$}}
\put(160,50){\makebox(0,0){$B_L$}}
\put(160,70){\makebox(0,0){$h$}}
\put(10,90){\makebox(0,0){Unbroken}}
\put(85,90){\makebox(0,0){Internal}}
\put(160,90){\makebox(0,0){Broken}}
\thicklines
\put(150,50){\vector(-1,0){30}}
\put(100,50){\vector(-1,0){30}}
\put(50,52.5){\vector(-2,1){30}}
\put(135,28){\makebox(0,0)[t]{$\scriptstyle P$}}
\put(135,52){\makebox(0,0)[b]{$\scriptstyle P$}}
\put(124,40){\makebox(0,0)[r]{$\scriptstyle Q$}}
\put(146,40){\makebox(0,0)[l]{$\scriptstyle Q$}}
\put(35,48){\makebox(0,0)[t]{$\half$}}
\put(35,74){\makebox(0,0)[b]{$\half$}}
\put(24,60){\makebox(0,0)[r]{$\half$}}
\put(46,60){\makebox(0,0)[l]{$\half$}}
\end{picture}
\caption{The unique path from $B_L$ to~$\ppl$.}
\label{f:connection-path}
\end{figure}

The results~\(asy-trans-prob) of the previous calculation
suggest a simpler method
which we describe for the case of transmission in the $\ppl$~mode.
As indicated in figure~\ref{f:connection-path}
there is only one path from $B_L$ to~$\ppl$.
Since there is no interference between different paths
we can calculate with probabilities instead of probability amplitudes
and can read off the transmission probability~$PT_\l/2$ directly.
Using this method we obtain most of the scattering probabilities
given in table~\ref{t:scattering}.
For generality we do not make use of the fact that $T_h \equiv 1$.
The only entries of table~\ref{t:scattering}
that require further calculation
are the ones that describe reflection of the modes $B_L$ and~$B_{T1}$
and reflection of the charged scalars $\ppl$~and~$\pmi$.
\begin{table}
$$ \begin{array}{|c|cccc|cccc|} \cline{2-9}
\mc&    \ppl&    \pmi&    A_{T1}&  A_{T2}& h&
  B_L&     B_{T1}&  B_{T2}\\ \cline{1-9}
\ppl&   R_\phi&  X_\phi&  \0&      \0&     T_h/2& PT_\l/2& QT_\l/2& \0\\
\pmi&   X_\phi&  R_\phi&  \0&      \0&     T_h/2& PT_\l/2& QT_\l/2& \0\\
A_{T1}& \0&      \0&      R_\tau&  \0&     \0&    QT_\tau& PT_\tau& \0\\
A_{T2}& \0&      \0&      \0&      R_\tau& \0&
  \0&      \0&      T_\tau\\ \cline{1-9}
h&      T_h/2&   T_h/2&   \0&      \0&     R_h&   \0&      \0&      \0\\
B_L&    PT_\l/2& PT_\l/2& QT_\tau& \0&     \0&    R_L&     X&       \0\\
B_{T1}& QT_\l/2& QT_\l/2& PT_\tau& \0&     \0&    X&       R_{T1}&  \0\\
B_{T2}& \0&      \0&      \0&      T_\tau& \0&
  \0&      \0&      R_\tau\\ \cline{1-9}
\end{array} $$
\caption{Scattering probabilities for asymptotic modes.
A blank entry indicates a probability of zero.}
\label{t:scattering}
\end{table}

\isubsection{Interference between internal modes}
\label{s:interfere}

We consider the reflection of an incident $B_L$~particle.
The amplitude vectors for the process are
\ble
\cvec{0\\
      1\vphantom{\xwk}\\
      0\vphantom{\xpM}\\
      0}
\via(inverted-broken-k)
\cvcC{0\\
      -\xwk\\
      \ghost\xpM\\
      0}
&\to&
\cvcC{0\\
      -r_\l\;\xwk\\
      \ghost r_\tau\;\xpM\\
      0} \\
&\via(connect-broken-k)&
\cvcc{0\\
      -r_\l\;\bigl(\xwk\bigr)^2 + r_\tau\;\bigl(\xpM\bigr)^2\\
      (r_\l+r_\tau)\;\xwk\;\xpM\\
      0}. \nonumber
\ele
We take the norm of the elements of the final amplitude vector
to obtain two reflection probabilities,
the probability~$R_L$ of reflection of~$B_L$ as itself
and the probability~$X$ of crossing, that is,
of reflection of~$B_L$ into~$B_{T1}$.
When we calculate the corresponding reflection probabilities
for a $B_{T1}$~particle,
we find that the crossing probability~$X$ also describes
reflection of~$B_{T1}$ into~$B_L$.
The reflection probabilities are
\begingroup
\def\hpl#1{\hphantom{PQ}\llap{$#1$}}
\be
R_L    &=& \hpl{P^2}\,R_\l + \hpl{Q^2}\,R_\tau
- 2PQ\sqrt{R_\l R_\tau}\>\chi \label{refl-prob-rl} \\
R_{T1} &=& \hpl{Q^2}\,R_\l + \hpl{P^2}\,R_\tau
- 2PQ\sqrt{R_\l R_\tau}\>\chi\\
X      &=&        PQ\,R_\l +        PQ\,R_\tau
+ 2PQ\sqrt{R_\l R_\tau}\>\chi, \label{refl-prob-x}
\ee
\endgroup
where the crossing enhancement factor~$\chi$ is defined by
\$ \chi = \re \biggl[ {r_\l^*r_\tau \over |r_\l||r_\tau|} \biggr]. \$

\begin{figure}\centered
\epsf{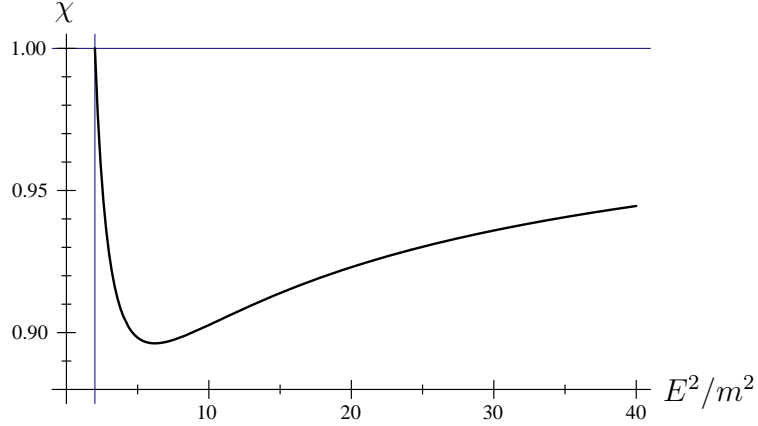}{2in}
\begin{picture}(0,0)
\put(22,10){\makebox(0,0){$E^2/m^2$}}
\put(-222,154){\makebox(0,0){$\chi$}}
\end{picture}
\caption{Crossing enhancement factor~$\chi$
as a function of~$E^2/m^2$ for a mass ratio~$\M^2$ of~2.}
\label{f:crossing-largeratio}
\end{figure}

The interference term which appears
in equations~\(refl-prob-rl)--\(refl-prob-x)
is proportional to the factor~$\chi$,
the cosine of the angle between the amplitudes $r_\l$~and~$r_\tau$.
When $\chi$~is positive,
the interference term increases the probability of crossing.
For mass ratios~$\M^2 > 1$
the behavior of~$\chi$ as a function of~$E^2/m^2$ is not complicated.
The curve shown in figure~\ref{f:crossing-largeratio} is typical---%
$\chi$ drops from the value~1 at the threshold~$E^2/m^2 = \M^2$
to a positive minimum,
then returns gradually to~1.

\begin{figure}\centered
\epsfx{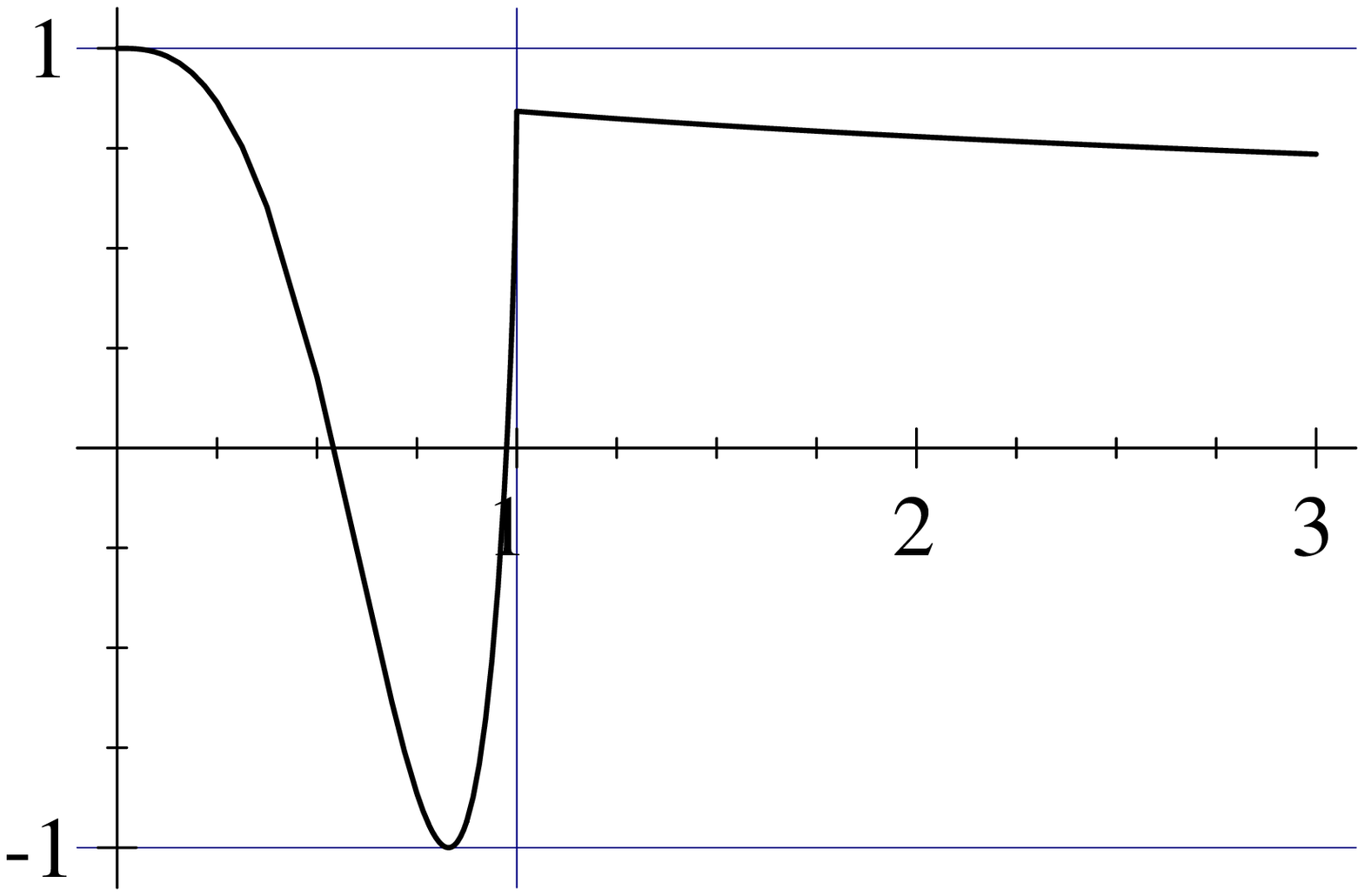}{1.6in}\hfill
\epsfx{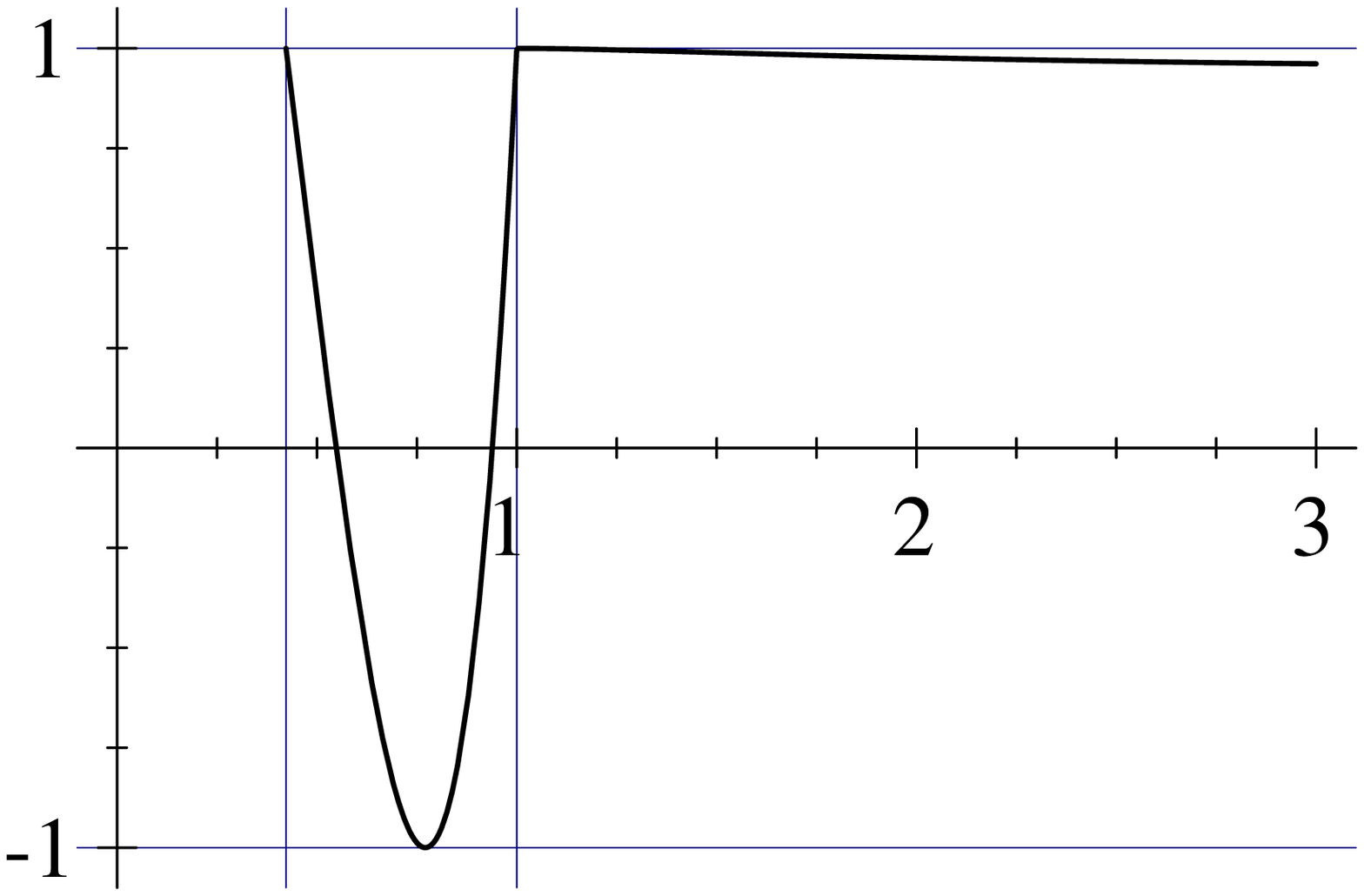}{1.6in}\hfill
\epsfx{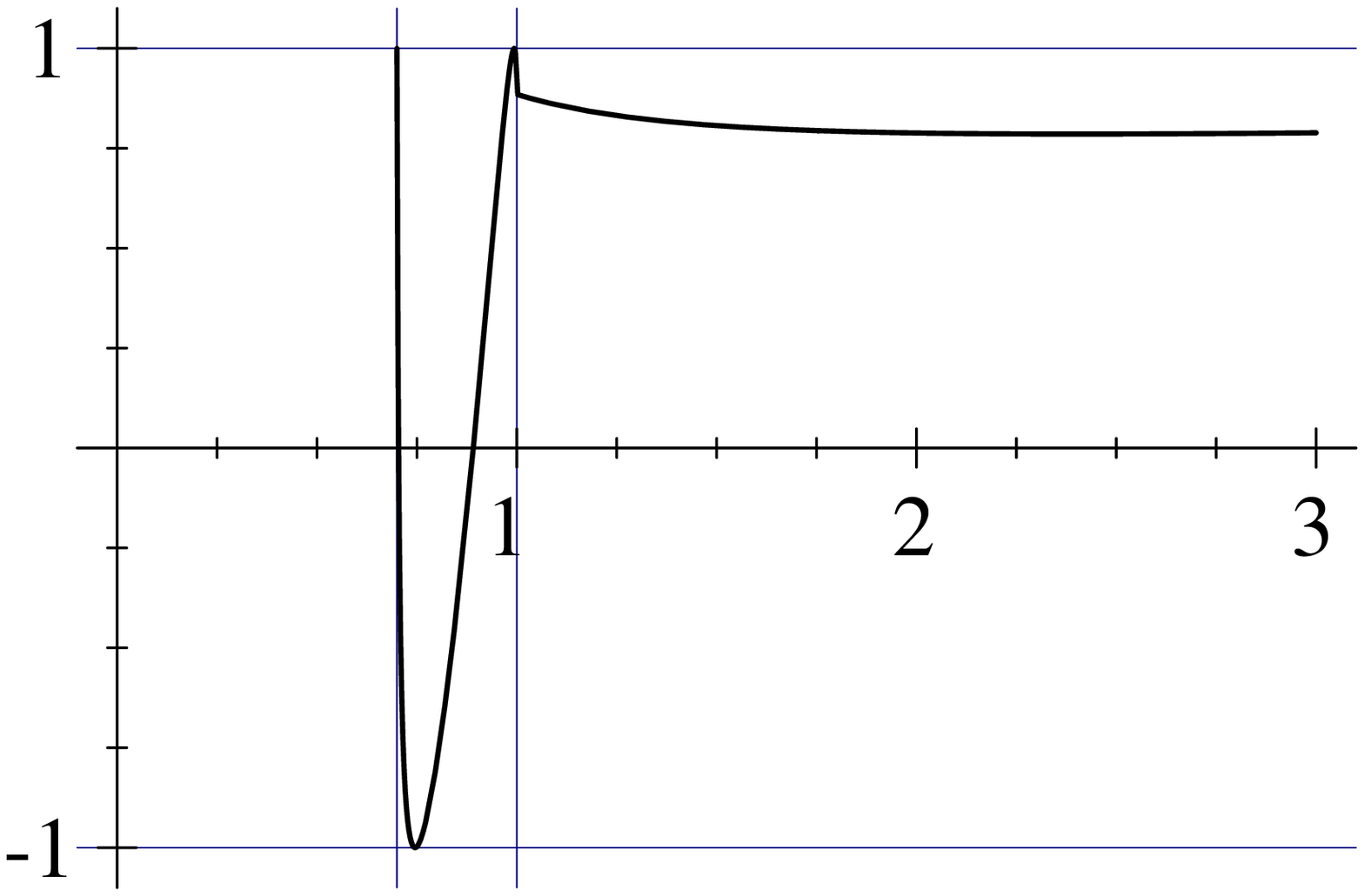}{1.6in}
\vskip18pt\leavevmode
\epsfx{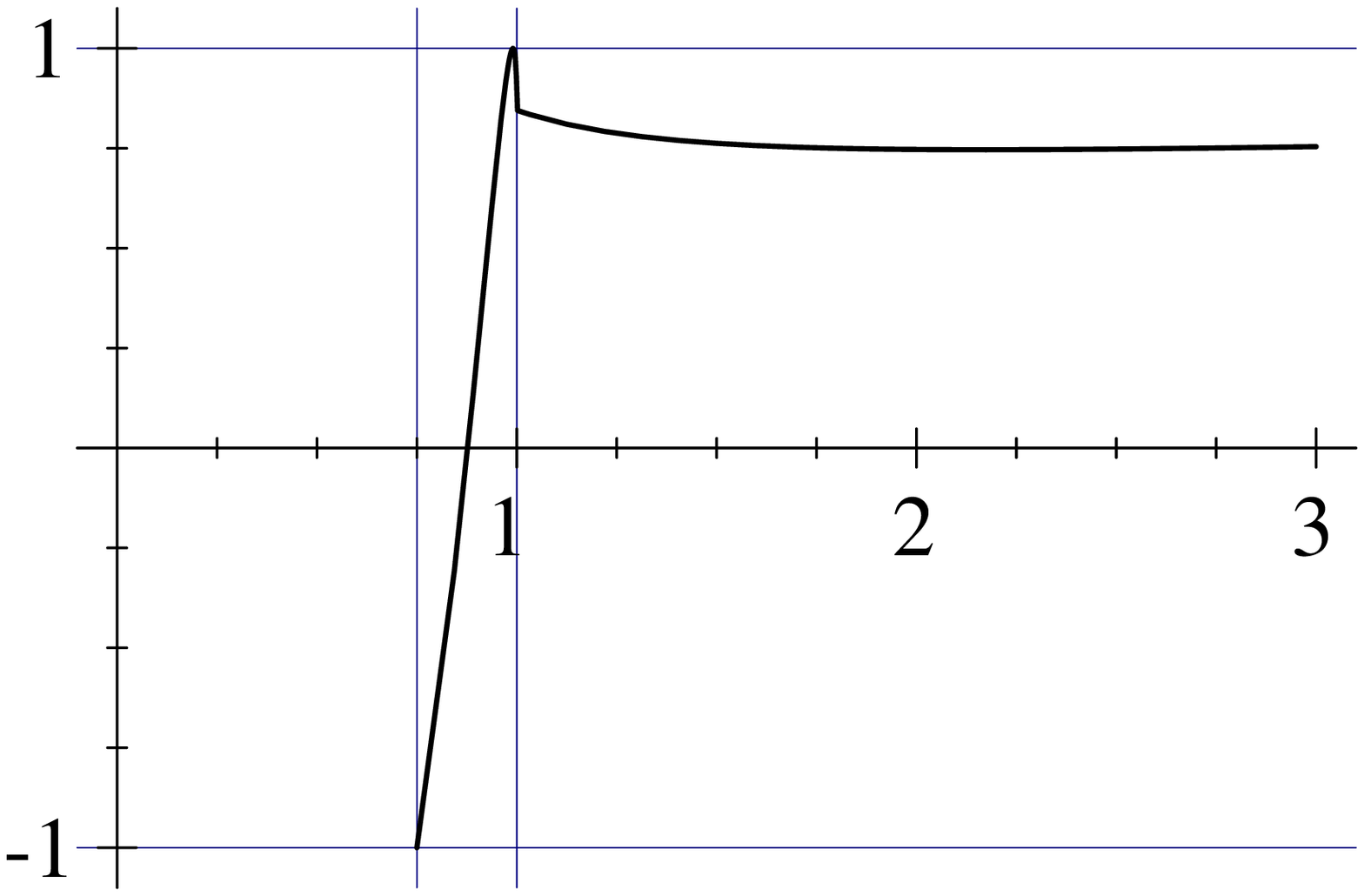}{1.6in}\hfill
\epsfx{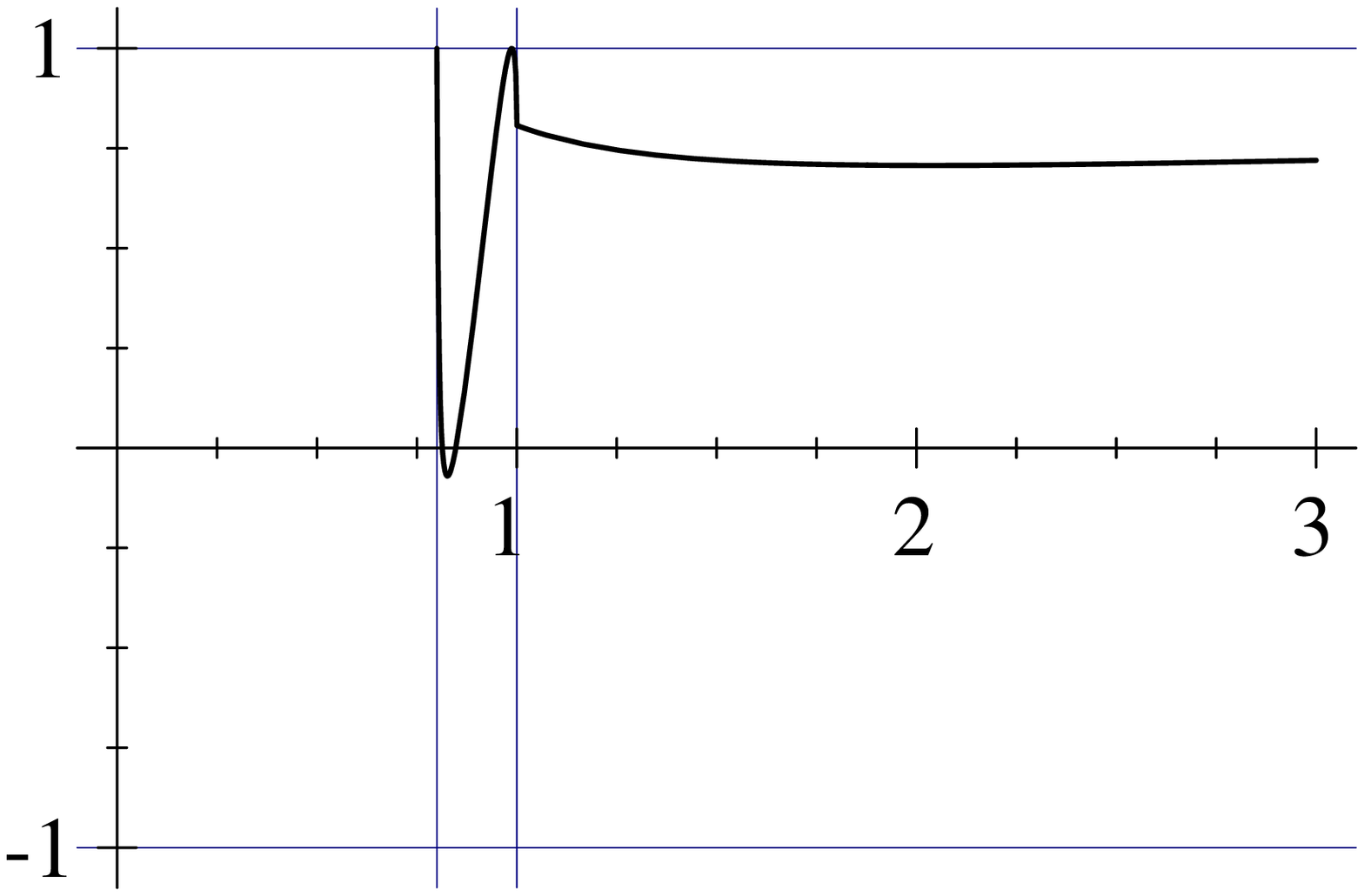}{1.6in}\hfill
\epsfx{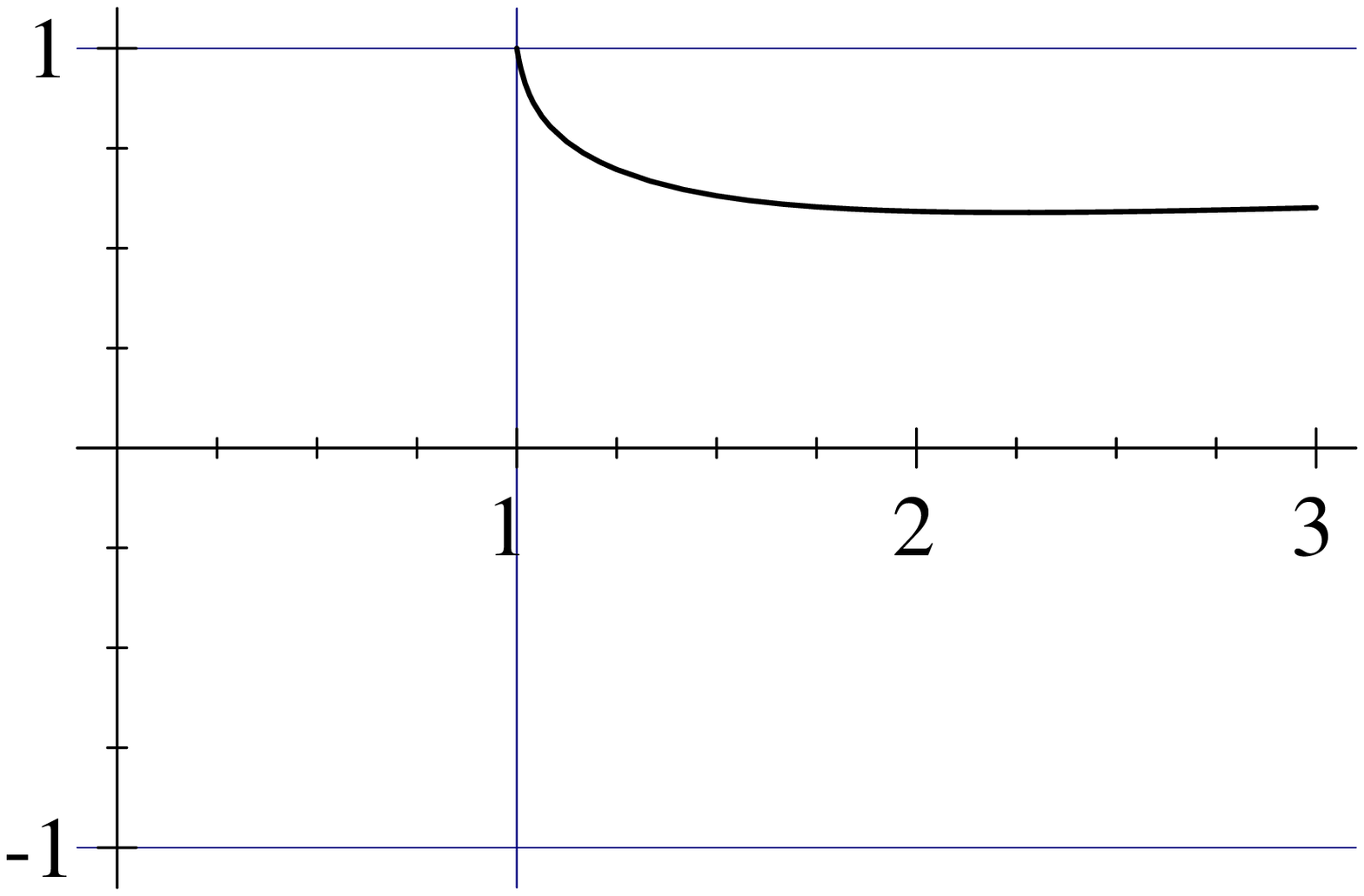}{1.6in}
\caption{Crossing enhancement factor~$\chi$ as a function of~$E^2/m^2$
for gauge boson mass ratios~$\M^2$ of 0, 0.4225, 0.7, 3/4, 0.8, and~1.}
\label{f:crossing}
\end{figure}

We plot a set of representative curves for mass ratios~$\M^2 < 1$
in figure~\ref{f:crossing}.
In the region~$E^2/m^2 > 1$ the curves all exhibit
the same positive minimum and gradual return to~1,
but in the region~$E^2/m^2 < 1$ where the $B_\l$~mode is totally reflected,
the curves are qualitatively different from one another.
The mass ratio~$\M^2$ at which
the cusp at~$E^2/m^2 = 1$ reaches unit height is only approximately~$0.4225$,
but the ratio at which the $\chi = -1$~minimum vanishes is {\em exactly}~3/4,
the same ratio at which the bound state in the $B_\l$~mode appears.

Analogous calculations for the reflection of
the charged scalars $\ppl$~and~$\pmi$
yield the reflection probabilities
\be
R_\phi &=& \textf1/4 \Bigl( R_h + R_\l
  - 2 \sqrt{R_h R_\l} \> \chi_\phi \Bigr) \\
X_\phi &=& \textf1/4 \Bigl( R_h + R_\l
  + 2 \sqrt{R_h R_\l} \> \chi_\phi \Bigr)
\ee
where
\$ \chi_\phi = \re \biggl[ \biggl( {k_m+im\over E}
  \biggr)^{\mkern-5mu 2} {r_h^*r_\l \over |r_h||r_\l|} \biggr]. \$
For the quartic Higgs potential~\(quartic)
these results for charged scalars sim\-pli\-fy---%
the amplitude~$r_h$ is identically zero,
$\chi_\phi$~is undefined,
and the reflection probabilities $R_\phi$ and~$X_\phi$
both reduce to~$R_\l/4$.

\isubsection{Discussion of results}
\label{s:discussion}

For particles with energy sufficiently far above all relevant mass thresholds
the internal mode transmission probabilities are essentially unity
and the process of scattering from the domain wall
can be regarded as a process of conversion of particles between phases.
In this limit the scattering probabilities in table~\ref{t:scattering}
reduce to the conversion probabilities in table~\ref{t:conversion}.
The breakdown of this approximation occurs when
either the momentum of the incident particle approaches zero
or one of the internal modes to which the particle
connects approaches total reflection.
\begin{table}
$$ \begin{array}{|c|cccc|} \cline{2-5}
\mc&    h&   B_L& B_{T1}& B_{T2}\\ \cline{1-5}
\ppl&   1/2& P/2& Q/2&    \0\\
\pmi&   1/2& P/2& Q/2&    \0\\
A_{T1}& \0&  Q&   P&      \0\\
A_{T2}& \0&  \0&  \0&     1\\ \cline{1-5}
\end{array} $$
\caption{Conversion probabilities for high-energy particles.}
\label{t:conversion}
\end{table}

We consider the extent to which
our results based on the quartic potential of equation~\(quartic)
apply when the Higgs potential is more general.
With the exception of the plots of~$\chi$,
the results of sections \ref{s:scatter-asymptotic}~and~\ref{s:interfere}
hold for an arbitrary Higgs potential---%
all dependence on the potential
is contained in the reflection and transmission amplitudes
for the internal modes.
{}From the discussion of bound states in section~\ref{s:bound-states}
we draw the following generally applicable conclusions:
The potential~$U_\tau$ is always monotone increasing,
so the $B_\tau$~modes never have bound states.
The $B_\l$~mode need not have a bound state,
but is capable of supporting at least one.
The $h$~mode is guaranteed to have at least one bound state.
We cannot make any more specific statement
about the bound states of the $h$~mode,
because for any positive integer there exist Higgs potentials~$V(v)$
for which the $h$~mode has that number of bound states.
Suitable Higgs potentials can be constructed
by joining three quadratic pieces;
the resulting scattering potentials~$U_h = V''(v)$
are square wells of various sizes.

\section{Extension to the standard model}
\label{sec:sm}
\isubsection{Reduction to two subproblems}

The fields appearing in the standard model are listed in table~\ref{t:fields},
along with their transformation properties under the various symmetry groups.
We take the symbols~$\Bmu$ and, later,~$\Amu$ to have
completely different meanings than in the previous sections.
\begin{table}
$$
\def\-{\hbox{-}}
\def\V{\surd}
\def\l#1{\llap{$#1$}}
\begin{array}{|c|ccccc|} \cline{2-6}
\mc&\hbox{Spin}&\hbox{Family}&Y&SU(2)&SU(3)\\ \cline{1-6}
B^\mu&    1&   \-& 0&      \-& \-\\
W^\mu&    1&   \-& 0&      \hbox{triplet}& \-\\
G^\mu&    1&   \-& 0&      \-& \hbox{octet}\\ \cline{1-6}
\phi&     0&   \-& \l+1&   \V& \-\\ \cline{1-6}
\ell_\sl& 1/2& \V& \l-1&   \V& \-\\
e_\sr&    1/2& \V& \l-2&   \-& \-\\ \cline{1-6}
q_\qsl&   1/2& \V& 1/3&    \V& \V\\
u_\sr&    1/2& \V& 4/3&    \-& \V\\
d_\sr&    1/2& \V& \l-2/3& \-& \V\\ \cline{1-6}
\end{array} $$
\caption{Fields appearing in the standard model.
A check indicates that the field transforms
according to the fundamental representation of the corresponding group.}
\label{t:fields}
\end{table}

The standard model Lagrangian is
\ble
\L \;\; = &-& \f1/4 B_{\mu\nu} B^{\mu\nu}
- \f1/4 W^i_{\mu\nu} W^{i\,\mu\nu}
- \f1/4 G^a_{\mu\nu} G^{a\,\mu\nu} \no \noalign{\medskip}
&+& (D_\mu \phi)^\dagger (D^\mu \phi) - V(\rt\,|\phi|)
\label{sm-lagrangian} \\ \noalign{\medskip}
&& \kern-5.3em {} + \biggl(\,
   \f{i}/2\bar\ell_\sl\Dslash\ell_\sl
 + \f{i}/2\bar e_\sr  \Dslash e_\sr
 + \f{i}/2\bar q_\qsl \Dslash q_\qsl
 + \f{i}/2\bar u_\sr  \Dslash u_\sr
 + \f{i}/2\bar d_\sr  \Dslash d_\sr \no
&& \kern-5.3em\kern8.25em {} - \bar\ell_\sl M_e \phi \mkern2mu e_\sr
 -  \bar q_\qsl  M_u \phi_c u_\sr
 -  \bar q_\qsl  M_d \phi \mkern2mu d_\sr \phc\biggr). \nonumber
\ele
For equation~\(sm-lagrangian) we define the norm~$|\phi|$ by
\$ |\phi| = \sqrt{\phi^\dagger \phi} \$
and the conjugate field~$\phi_c$ by
\$ \phi_c = i\tau_2\phi^*. \$
The constants~$M_e$, $M_u$, and~$M_d$ are matrices in family space.

To study scattering from a domain wall at the electroweak phase transition,
we will write out the components of~$\phi$,
\$ \phi = \f1/\rt \biggl( {\p3+i\p4 \atop \p1+i\p2} \biggr),
\label{phi-components} \$
derive equations of motion,
define
$$ \p1 = v + h, \eqno{\(p1-expansion)} $$
take all fields except~$v$ to be perturbatively small,
and obtain an equation for the static domain wall solution~$v$
along with first-order equations of motion
which describe perturbations about it.
First, however,
we simplify the calculation by removing from the Lagrangian
terms which do not contribute to the end result,
such as terms higher than quadratic in fields other than~$\phi$.

Specifically,
we remove the bilinear terms from the gauge field strengths
and the gauge couplings from the fermion covariant derivatives.
The fermion mass terms are quadratic
in the perturbatively small fermion fields
and so do not contribute to the equations of motion for~$\phi$.
Consequently, we can expand~$\phi$ according to
equations \(phi-components)~and~\(p1-expansion)
even before deriving the equations of motion.
Dropping higher than quadratic terms yields the simplified mass terms
\$ \biggl(
 - \bar e_\sl {vM_e\over\rt} e_\sr
 - \bar u_\sl {vM_u\over\rt} u_\sr
 - \bar d_\sl {vM_d\over\rt} d_\sr \phc\biggr). \$
Redefining the fermion fields by applying
independent unitary rotations in family space
to the left and right components of the fields $e$, $u$, and~$d$
allows us to take the mass matrices~$M_e$, $M_u$,
and~$M_d$ to be diagonal with positive real eigenvalues.

As a result of the above transformations
the Lagrangian~\(sm-lagrangian) falls into unrelated pieces.
For each massive fermion species there is a copy of the Dirac Lagrangian
\$ \biggl(\f{i}/2 \bar\psi \dslash\psi \phc\biggr)
- (\mu v) \bar\psi \psi \label{lagr-fermion} \$
with a broken-phase mass~$\mu\vo$ equal to
$\vo/\rt$~times the eigenvalue of the appropriate mass matrix.
We consider the massive fermions further in section~\ref{s:fermions}.
The pieces of the Lagrangian which describe neutrinos
\$ \biggl(\f{i}/2 \bar\nu_\sl \dslash \nu_\sl \phc\biggr) \$
and gluons
\$ - \f1/4 (\d_\mu G^a_\nu - \d_\nu G^a_\mu)(\d^\mu G^{a\,\nu}
- \d^\nu G^{a\,\mu}) \$
are Lagrangians of free particles.
To first order the neutrinos and gluons do not interact with the domain wall,
and we do not consider them further.
The remainder of the original Lagrangian,
\ble
- \f1/4 B_{\mu\nu} B^{\mu\nu} &-& \f1/4 (\d_\mu W^i_\nu
- \d_\nu W^i_\mu)(\d^\mu W^{i\,\nu} - \d^\nu W^{i\,\mu}) \no
&+& \hphantom{\f1/4} (D_\mu \phi)^\dagger (D^\mu \phi)
- V(\rt\,|\phi|), \label{lagr-boson}
\ele
we analyze in the following section.

\isubsection{Standard model bosons}

Having reduced the standard model Lagrangian~\(sm-lagrangian)
to the boson Lagrangian~\(lagr-boson),
we apply the method described in the previous section
and derive equations of motion.
After some calculation,
we find that the condition for the field configuration~$v(z)$ to be stable
$$ v'' = V'(v) \eqno{\(eom-v)} $$
and the first-order equation of motion for the field~$h$
$$ \dbox h = - V''(v) h \eqno{\(eom-h)} $$
are the same as in the Abelian Higgs model.
The equations of motion for the fields $\p3$ and~$W^{2\,\mu}$,
\be
\dbox\p3 & = & - \textf1/2 gv\,\d\cdot W^2 - g W^{2\,\mu} \d_\mu v
- {V'(v)\over v}\p3 \\
\d_\nu \Wtilde^{2\,\mu\nu} &=& \ghost\textf1/2 g(v\d^\mu\p3
- \p3\d^\mu v) + (\textf1/2 gv)^2 W^{2\,\mu},
\ee
where
\$ \Wtilde^{i\,\mu\nu} = \d^\mu W^{i\,\nu} - \d^\nu W^{i\,\mu}, \$
are exact analogues of equations \(eom-p2) and~\(eom-A) for $\p2$ and~$\Amu$,
as are the equations of motion for $\p4$ and~$W^{1\,\mu}$:
\be
\dbox\p4 & = & - \textf1/2 gv\,\d\cdot W^1 - g W^{1\,\mu} \d_\mu v
- {V'(v)\over v}\p4 \\
\d_\nu \Wtilde^{1\,\mu\nu} &=& \ghost\textf1/2 g(v\d^\mu\p4
- \p4\d^\mu v) + (\textf1/2 gv)^2 W^{1\,\mu}.
\ee

To cast the last three equations of motion
\begingroup
\def\'{\hphantom{'}}
\be
\dbox\p2 & = & - \textf1/2  v\,\d\cdot(g'B-gW^3)
- (g'B^\mu-gW^{3\,\mu}) \d_\mu v - {V'(v)\over v}\p2 \qquad \\
\d_\nu \Wtilde^{3\,\mu\nu} &=&      -\textf1/2 g\'(v\d^\mu\p2
- \p2\d^\mu v) - \textf1/4  g\'v^2(g'B^\mu-gW^{3\,\mu})
   \vrule depth 9pt width 0pt \\
\d_\nu B^{\mu\nu}          &=& \ghost\textf1/2 g' (v\d^\mu\p2
- \p2\d^\mu v) + \textf1/4  g' v^2(g'B^\mu-gW^{3\,\mu})
\ee
\endgroup
into a form analogous to equations \(eom-p2) and~\(eom-A),
we define the linear combinations
\$
Z^\mu = {g'B^\mu-g W^{3\,\mu} \over \gsqr} \_and_
A^\mu = {g B^\mu+g'W^{3\,\mu} \over \gsqr}
\$
and obtain equations of motion for $\p2$ and~$Z^\mu$
\be
\dbox\p2 &=& - \textf1/2 \gsqr\;v\;\d\cdot Z
- \gsqr\; Z^\mu \d_\mu v - {V'(v)\over v}\p2 \\
\d_\nu Z^{\mu\nu} &=& \ghost \textf1/2 \gsqr\, (v\d^\mu\p2
- \p2\d^\mu v)\, +\, (\textf1/2\gsqr\,v)^2 \, Z^\mu
\ee
along with an equation of motion
\$ \d_\nu F^{\mu\nu} = 0 \$
which describes a noninteracting massless gauge boson---the photon.

We summarize the correspondence
between the standard model and the Abelian Higgs model
in the following table.
\$ \begin{array}{|lcc|ll|} \cline{1-5}
\Amu&\p2&e&\Bmu&M\\ \cline{1-5}
\openabove{W^{1\,\mu}}&\p4&\textf1/2g&\W^{1\,\mu}&M_W\\
\openabove{W^{2\,\mu}}&\p3&\textf1/2g&\W^{2\,\mu}&M_W\\
\openabove{Z^\mu}&\p2&\openbelow{\textf1/2\gsqr}&\Z^\mu&M_Z\\ \cline{1-5}
\end{array} \$
The right side of the table shows the names we assign
to the analogues of~$\Bmu$ and~$M$.
{}From the defining equations \(define-B) and~\(mass-gauge-boson)
for $\Bmu$~and~$M$ we find,
for example,
\be
\W^{2\,\mu} &=& W^{2\,\mu} + \d^\mu \biggl( {\p3 \over \f1/2 gv} \biggr) \\
M_W^2 &=& (\textf1/2 g\vo)^2.
\ee

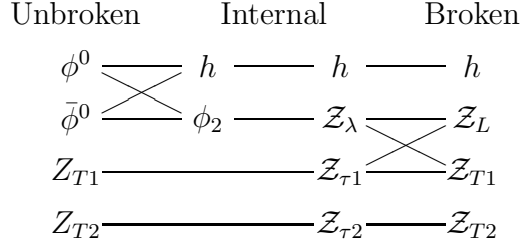
\begin{figure}\centered
\begin{picture}(170,100)
\thinlines
\multiput(20,10)(0,20){2}{\line(1,0){80}}
\multiput(20,50)(0,20){2}{\line(1,0){30}}
\multiput(70,50)(0,20){2}{\line(1,0){30}}
\multiput(120,10)(0,20){4}{\line(1,0){30}}
\put(20,52.5){\line(2,1){30}}
\put(20,67.5){\line(2,-1){30}}
\put(120,32.5){\line(2,1){30}}
\put(120,47.5){\line(2,-1){30}}
\put(10,10){\makebox(0,0){$Z_{T2}$}}
\put(10,30){\makebox(0,0){$Z_{T1}$}}
\put(10,50){\makebox(0,0){$\bar\phi^0$}}
\put(10,70){\makebox(0,0){$\phi^0$}}
\put(60,50){\makebox(0,0){$\p2$}}
\put(60,70){\makebox(0,0){$h$}}
\put(110,10){\makebox(0,0){$\Z_{\tau2}$}}
\put(110,30){\makebox(0,0){$\Z_{\tau1}$}}
\put(110,50){\makebox(0,0){$\Z_\l$}}
\put(110,70){\makebox(0,0){$h$}}
\put(160,10){\makebox(0,0){$\Z_{T2}$}}
\put(160,30){\makebox(0,0){$\Z_{T1}$}}
\put(160,50){\makebox(0,0){$\Z_L$}}
\put(160,70){\makebox(0,0){$h$}}
\put(10,90){\makebox(0,0){Unbroken}}
\put(85,90){\makebox(0,0){Internal}}
\put(160,90){\makebox(0,0){Broken}}
\end{picture}
\caption{Connections between neutral modes in the standard model.}
\label{f:connection-neutral}
\end{figure}

By applying the above correspondence
to our results for the Abelian Higgs model
we can find out nearly everything we want to know about standard model bosons.
To be specific, we obtain
the definitions and scattering potentials of the scalar internal modes
and most of the equations that connect
the internal modes to the asymptotic ones.
What the correspondence fails to provide
is the connection between the real Higgs modes $h$,~$\p2$, $\p3$ and~$\p4$
and the complex modes appropriate to the unbroken phase.
The modes $h$~and~$\p2$ combine just as in the Abelian Higgs model---%
copying equation~\(particle-antiparticle)
we define the amplitudes of the particle and antiparticle solutions
\$ \phi^0 = {h+i\p2\over\rt} \_and_ \bar\phi^0 = {h-i\p2\over\rt} \$
and obtain the connection diagram given in figure~\ref{f:connection-neutral}.
The associated connection equations,
scattering probabilities, and conversion probabilities
may be found by applying the correspondence to
equations \(connect-broken)~and~\(connect-unbroken)
and tables \ref{t:scattering}~and~\ref{t:conversion}.

Since the Higgs fields $\p3$~and~$\p4$ combine with each other
rather than with the field~$h$
in the charged particle and antiparticle amplitudes
\$ \ppl = {\p3+i\p4\over\rt} \_and_ \pmi = {\p3-i\p4\over\rt},
   \label{phi-rotation} \$
the connection equations involving these modes
will not be exact analogues of the connection equations
of the Abelian Higgs model.
Taking the relevant parts of equations
\(connect-broken)~and~\(connect-unbroken)
and applying the correspondence,
we obtain the reduced connection equations
\begingroup
\def\vph{\vphantom{\pM}}
\$
\cvec{\p3\vph\\W^2_{T1}\vph\\W^2_{T2}} =
\left(\begin{array}{ccc}
  {m+ik\over E}\vph&\0&\0\\
  \0\vph&\sgn k&\0\\
  \0&\0&1\\
\end{array}\right)
\cvec{\W^2_\l\vph\\ \W^2_{\tau1}\vph\\ \W^2_{\tau2}}
\label{sm-connect-unbroken}
\$
and
\$
\cvec{\W^2_L\vph\\ \W^2_{T1}\vph\\ \W^2_{T2}} =
\left(\begin{array}{ccc}
  \ghost\wk&\pM&\0\\
  -\pM&\wk&\0\\
  \0&\0&1\\
\end{array}\right)
\cvec{\W^2_\l\vph\\ \W^2_{\tau1}\vph\\ \W^2_{\tau2}}
\label{sm-connect-broken}
\$
\endgroup
along with a similar pair for the fields $W^1$~and~$\p4$.
We define charged gauge boson amplitudes
for each of the internal and asymptotic gauge boson modes,
for example,
\$ W^+_{T1}  = {W^2_{T1}+iW^1_{T1}\over\rt}
\_and_ W^-_{T1} = {W^2_{T1}-iW^1_{T1}\over\rt}. \label{W-rotation} \$
Since the definitions \(phi-rotation)~and~\(W-rotation)
describe the same linear transformation of amplitudes,
the connection matrices for the charged bosons
are identical to those in equations
\(sm-connect-unbroken)~and~\(sm-connect-broken).
We present one of the associated connection diagrams
in figure~\ref{f:connection-charged};
the absence of connections between positive and negative bosons
is of course guaranteed by the conservation of electric charge.\footnote{%
  The correct analogue of the charge non-conservation
  in scattering in the Abelian Higgs model
  is the non-conservation of isospin and hypercharge,
  for example, in the reflection process~$\phi^0 \to \bar\phi^0$.}
Applying the methods
of sections \ref{s:scatter-asymptotic}~and~\ref{s:interfere}
produces the scattering probabilities
given in table~\ref{t:scattering-charged}
and the conversion probabilities,
that is, the transmission probabilities for high-energy particles,
given in table~\ref{t:conversion-charged}.
\begin{figure}\centered
\begin{picture}(170,80)
\thinlines
\multiput(25,10)(0,20){3}{\line(1,0){60}}
\multiput(115,10)(0,20){3}{\line(1,0){30}}
\put(115,32.5){\line(2,1){30}}
\put(115,47.5){\line(2,-1){30}}
\put(10,10){\makebox(0,0){$W^+_{T2}$}}
\put(10,30){\makebox(0,0){$W^+_{T1}$}}
\put(10,50){\makebox(0,0){$\ppl$}}
\put(100,10){\makebox(0,0){$\W^+_{\tau2}$}}
\put(100,30){\makebox(0,0){$\W^+_{\tau1}$}}
\put(100,50){\makebox(0,0){$\W^+_\l$}}
\put(160,10){\makebox(0,0){$\W^+_{T2}$}}
\put(160,30){\makebox(0,0){$\W^+_{T1}$}}
\put(160,50){\makebox(0,0){$\W^+_L$}}
\put(10,70){\makebox(0,0){Unbroken}}
\put(85,70){\makebox(0,0){Internal}}
\put(160,70){\makebox(0,0){Broken}}
\end{picture}
\caption{Connections between charged modes in the standard model.}
\label{f:connection-charged}
\end{figure}
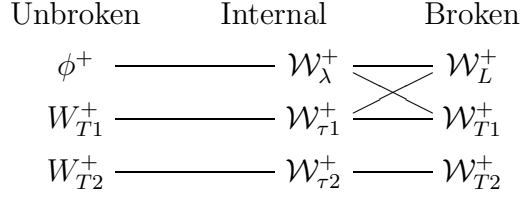
\begin{table}
$$ \begin{array}{|c|ccc|ccc|} \cline{2-7}
\mc&   \ppl&  W^+_{T1}& W^+_{T2}& \W^+_L& \W^+_{T1}& \W^+_{T2}\\ \cline{1-7}
\ppl&      R_\l&  \0&       \0&       PT_\l&   QT_\l&     \0\\
W^+_{T1}&  \0&    R_\tau&   \0&       QT_\tau& PT_\tau&   \0\\
W^+_{T2}&  \0&    \0&       R_\tau&   \0&      \0&        T_\tau\\ \cline{1-7}
\W^+_L&    PT_\l& QT_\tau&  \0&       R_L&     X&         \0\\
\W^+_{T1}& QT_\l& PT_\tau&  \0&       X&       R_{T1}&    \0\\
\W^+_{T2}& \0&    \0&       T_\tau&   \0&      \0&        R_\tau\\ \cline{1-7}
\end{array} $$
\caption{Scattering probabilities for charged modes in the standard model.}
\label{t:scattering-charged}
\end{table}
\begin{table}
$$ \begin{array}{|c|ccc|} \cline{2-4}
\mc&      \W^+_L& \W^+_{T1}& \W^+_{T2}\\ \cline{1-4}
\ppl&     P&      Q&         \0\\
W^+_{T1}& Q&      P&         \0\\
W^+_{T2}& \0&     \0&        1\\ \cline{1-4}
\end{array} $$
\caption{Conversion probabilities for charged modes in the standard model.}
\label{t:conversion-charged}
\end{table}

\isubsection{Scattering of fermions}
\label{s:fermions}

Variation of the Dirac Lagrangian~\(lagr-fermion) gives the Dirac equation
\$ (i\dslash - \mu v)\psi = 0. \label{eom-psi} \$
To reduce this equation to scattering equations for scalar internal modes
we use the ansatz of Ayala et al~\cite{ayala}
\$ \psi = (i\dslash + \mu v)\tilde\psi \label{define-tildepsi} \$
to obtain
\$ [ \dbox + (\mu v)^2 - i\gamma^3(\mu v') ]\,\tilde\psi = 0. \$
We take $\tilde\psi$~to be proportional to
an eigenvector~$u_\pm$ of the matrix~$-i\gamma^3$ with eigenvalue~$\pm 1$,
\$ \tilde\psi = u_\pm\,f_\pm, \label{define-uf} \$
and find that the corresponding scalar function~$f_\pm$
satisfies the scattering equation
\$ [ \dbox + U_{f\pm}(z) ] \, f_\pm = 0 \label{eom-f} \$
with potential
\$ U_{f\pm} = (\mu v)^2 \pm \mu v'. \$

The existence of two different potentials for fermion scattering
is an artifact of the process of extracting scalar modes,
because any solution~$\psi$ of equation~\(eom-psi)
can be expressed in terms of either $f_+$ or~$f_-$.
For example,
the solution~$\psi$ generated by an eigenvector~$u_-$ and function~$f_-$
can also be obtained from the eigenvector~$u_+$ and function~$f_+$ given by
\be
u_+ &=& \f1/E \biggl( \,\sum_{\mu\ne 3} \gamma^\mu p_\mu \biggr) u_- \\
f_+ &=& \f1/E \biggl( \mu v + \d_3 \biggr) f_-.
\ee
We will express our results in terms of $u_+$ and~$f_+$,
because of the following advantage of the potential~$U_{f+}$:
being the sum of two positive terms which go to zero at~$\minf$,
it has no absolute minimum and so clearly has no bound states.

Although we have reduced the Dirac equation~\(eom-psi)
to a scalar equation~\(eom-f),
we have not yet defined internal and asymptotic modes for the fermion field.
For the moment we consider only positive-energy solutions
of equation~\(eom-psi).
In the Pauli-Dirac representation of the matrices~$\gamma^\mu$,
the normalized asymptotic solutions for the field~$\psi$
can be written in terms of a two-component spinor~$\varphi$ as
\$ \psi[\varphi] = {1\over 2\sqrt{\w+\mu v}}
   \bvec{\w+\mu v \atop \skew{-2}\vec p\cdot\skew{-2}\vec\sigma}
\, \varphi \; \eipx. \$
Our normalization---%
setting the $z$-component of energy flux equal to~$\f1/2 \w k$---%
is compatible with massless particles,
so these asymptotic solutions can be used
in both the broken and unbroken phases.
The two-component spinor~$\varphi$ describes
the spin of the fermion in its rest frame;
our normalization assumes it satisfies
\$ \varphi^\dagger \varphi = 1. \$

To study the relation between internal and asymptotic solutions
we consider a solution~$f_+$ which is asymptotically
a unit-amplitude plane wave:
\$ f_+ = \eipx. \$
We write the eigenvector~$u_+$ in the form
\$ u_+ = (1-i\gamma^3)\bvec{\chi\atop 0}, \$
where $\chi$~is a constant two-component spinor.
Substituting into the definitions \(define-tildepsi)~and~\(define-uf),
we find that the resulting asymptotic field~$\psi$ is equal to $\psi[\varphi]$
for the following unnormalized spinor~$\varphi$:
\$ \varphi = {2\over\sqrt{\w+\mu v}}
\left(\begin{array}{cc} \w+\mu v-ik&ip\\-ip&\w+\mu v-ik\\
\end{array}\right) \chi. \label{asymptotic-fermion} \$
As before, we choose the coordinate system so that
the transverse momentum~$p$ lies along the $x$-axis.
Because the values of the quantities $k$~and~$v$
appearing in equation~\(asymptotic-fermion)
depend on whether the particle is in the broken or unbroken phase
and on what the sign of the $z$-momentum is,
the spin direction determined by~$\varphi$
can change as a result of the scattering process.
However,
the particular choices
\$ \chi = \bvec{1\atop\pm i} \$
generate spinors
\$ \varphi = {2\over\sqrt{\w+\mu v}} (\w+\mu v\mp p -ik) \> \chi
   \label{unnormalized-phi} \$
which point in the directions $\hat y$~and~$-\hat y$
independent of the values of $k$~and~$v$.
Accordingly,
we define the internal modes $f_{+\up}$~and~$f_{+\down}$
to have spins aligned with the $y$-axis,
that is, to have the normalized polarization spinors
\$ \chi_\up = {1\over 2\sqrt{2(\w-p)}} \>\> \f1/\rt \bvec{1\atop  i} \_and_
\chi_\down  = {1\over 2\sqrt{2(\w+p)}} \>\> \f1/\rt \bvec{1\atop -i}. \$
Although equation~\(unnormalized-phi) contains
the nonconstant quantities $k$~and~$v$,
the normalized spinors $\chi_\up$~and~$\chi_\down$ are constant, as required,
as a result of the identity
\$ |\w+\mu v\mp p-ik|^2 = 2(\w+\mu v)(\w\mp p). \$

Defining the asymptotic modes
\$ \psi_\up = \psi\Biggl[\f1/\rt\bvec{1\atop  i}\Biggr] \_and_
\psi_\down  = \psi\Biggl[\f1/\rt\bvec{1\atop -i}\Biggr] \$
we obtain the connection equation
\$
\def\phase#1{{\w+\mu v #1 p-ik\over \sqrt{(\w+\mu v #1 p)^2 + k^2}}}
\def\vph{\vphantom{\phase-}}
\def\drop#1{\vphantom{f}\smash{\lower4pt\hbox{$#1$}}}
\cvec{\psi_\up\vph\\ \drop{\psi_\down}\vph} =
\left(\begin{array}{cc} \phase-&\0 \\ \0&\phase+ \\ \end{array}\right)
\cvec{f_{+\up}\vph\\ \drop{f_{+\down}}\vph} \label{connect-fermion}
\$
which is valid in both the broken and unbroken phases
when the appropriate values of $k$~and~$v$ are used.
Because the connection matrix is diagonal,
we conclude that our chosen asymptotic modes do not interconnect
and that the scattering probabilities for each of the asymptotic modes
are equal to the scattering probabilities $R_{f+}$~and~$T_{f+}$
for the internal mode~$f_+$.
The appearance of nontrivial complex phases in
the connection matrix of equation~\(connect-fermion)
indicates that in general the spin of the fermion
rotates about the $y$-axis during scattering.

To obtain the negative-energy solutions of the Dirac equation~\(eom-psi)
we make use of the charge conjugation symmetry\footnote{%
  We do not simply take~$\w$ to be negative,
  because the analysis of asymptotic behavior in section~\ref{s:solution}
  relies on the positive energy of the solution.}
\$ \psi \to i\gamma^2 \psi^*. \$
Applying charge conjugation to an incident negative-energy particle
produces an incident positive-energy particle,
for which we know the solution of equation~\(eom-psi).
Applying charge conjugation again
yields the desired negative-energy solution.
Under charge conjugation the magnitudes of
the coefficients of the scattered waves do not change,
so the negative-energy solutions have
the same reflection and transmission probabilities $R_{f+}$~and~$T_{f+}$
as the positive-energy solutions.

For the quartic Higgs potential~\(quartic),
the scattering potential~$U_{f+}$ can be written
in the standard form~\(standard-form),
so we can obtain the reflection and transmission probabilities
$R_{f+}$~and~$T_{f+}$
from equations \(refl-prob)~and~\(trans-prob).
For reference, the coefficients of the potential~$U_{f+}$ are
\be
U_0 &=& 0 \\
U_1 &=& \Mf^2 \\
U_2 &=& \Mf(1-\Mf),
\ee
where the mass ratio~$\Mf^2$ is defined to be $(\mu\vo)^2/m^2$.
The scattering probabilities so obtained
agree with those of Ayala et al~\cite{ayala}.
Our discussion of fermion scattering differs from theirs primarily in that
we have explicitly considered the polarization of the fermion.

\begin{figure}\centered
\epsfx{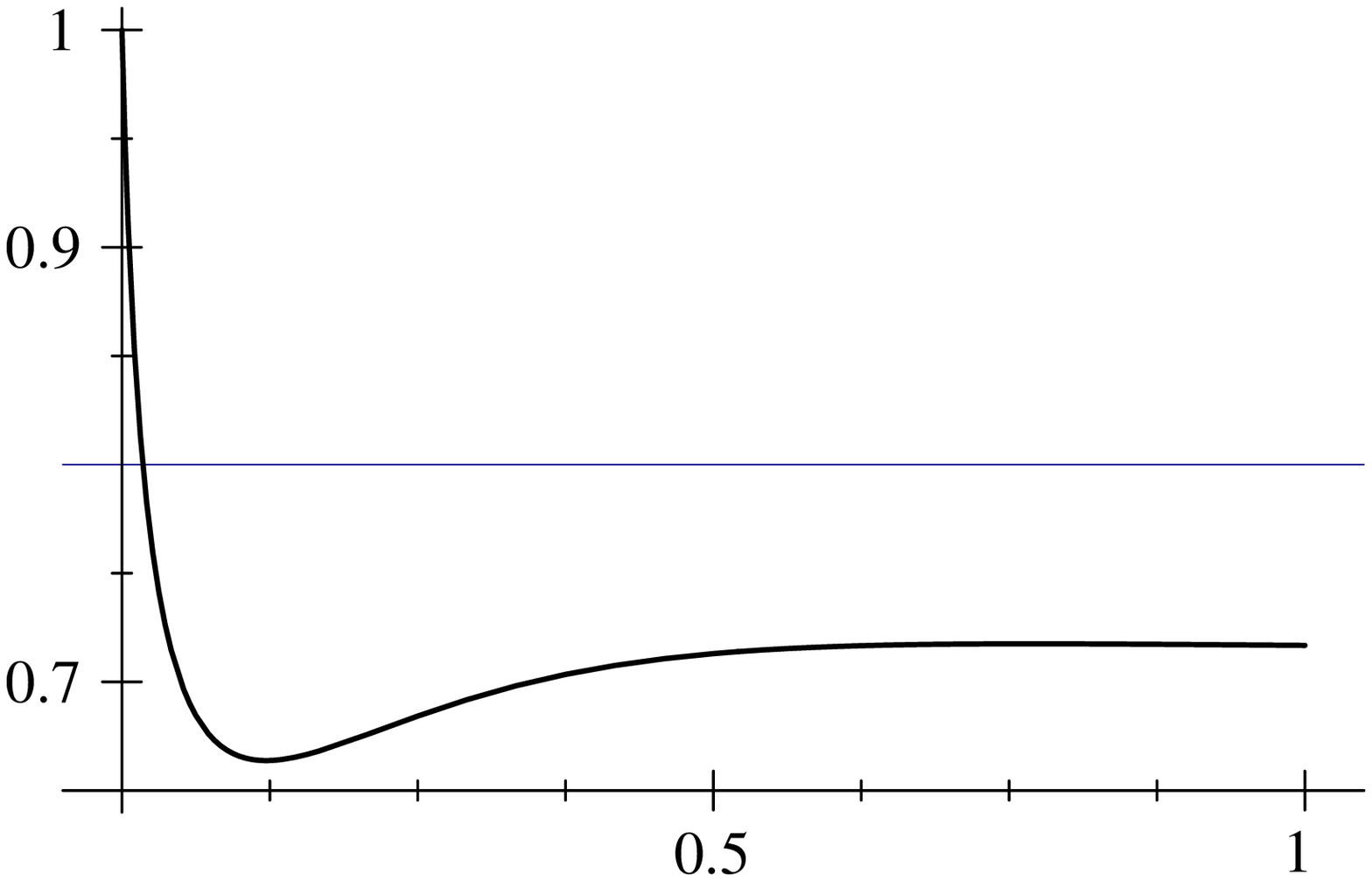}{2.4in}
\begin{picture}(0,0)
\put(13,9){\makebox(0,0){$\Mf^2$}}
\put(-158,118){\makebox(0,0){$T$}}
\end{picture}
\hfill
\epsfx{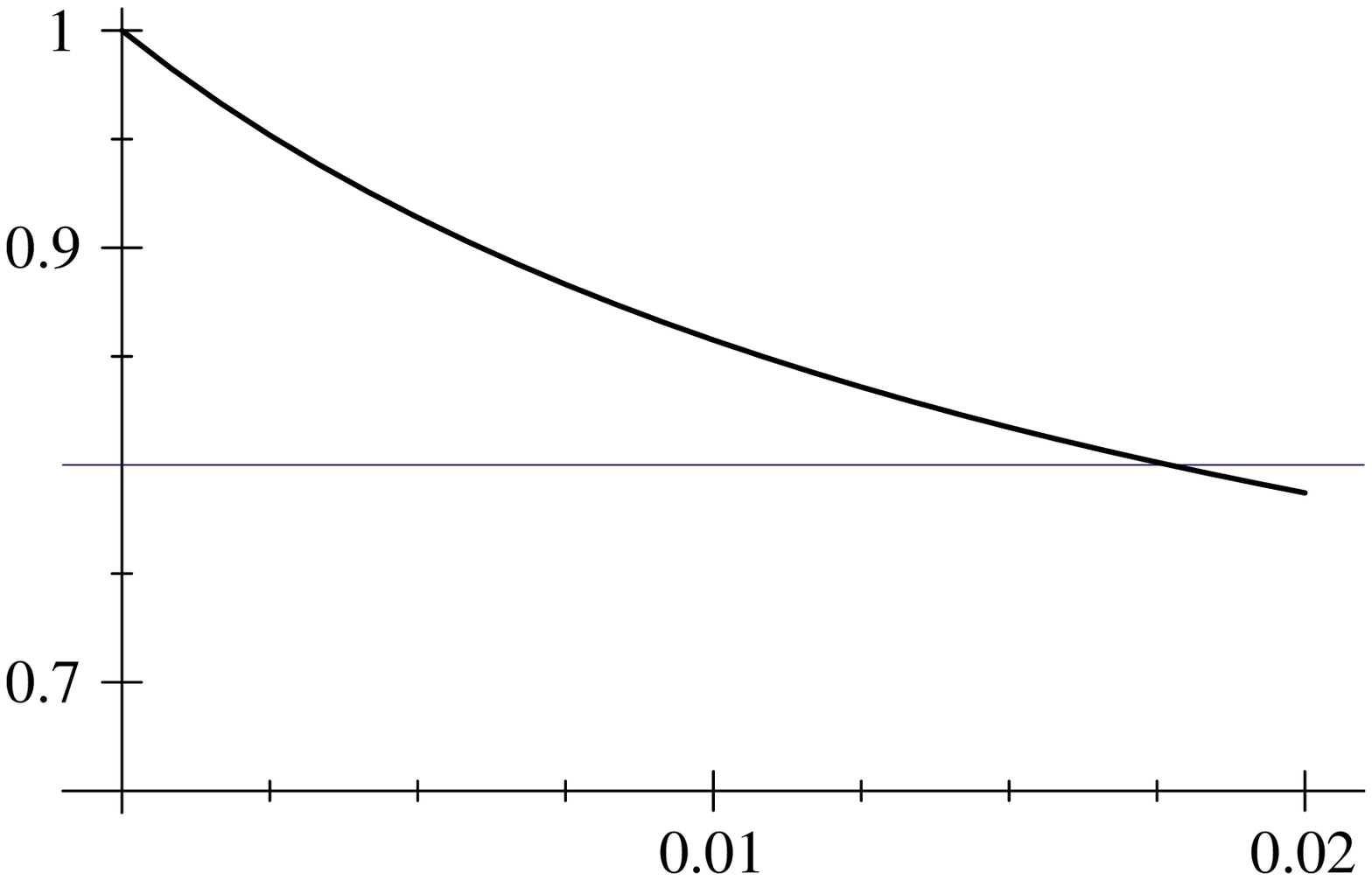}{2.4in}
\begin{picture}(0,0)
\put(13,9){\makebox(0,0){$\Mf^2$}}
\put(-158,118){\makebox(0,0){$T$}}
\end{picture}
\caption{Transmission probability near threshold for the $f_+$~mode,
plotted as a function of~$\Mf^2$.
The energy $E^2$ is taken to be $(\mu\vo)^2 + 0.01\,m^2$.}
\label{f:sharp}
\end{figure}

According to the results of section~\ref{s:scatter-internal}
the transmission probability as a function of~$E^2/m^2$
has the generic form described in that section
unless the scattering potential
is close to a potential which gives $T \equiv 1$,
in which case the onset of transmission is more abrupt.
Just such an increasing abruptness of onset
occurs in the $f_+$~mode
as the mass ratio~$\Mf^2$ goes to zero.
The effect can be seen in figure~\ref{f:sharp},
in which we plot the transmission probability
at a fixed distance above the threshold of total reflection
as a function of the mass ratio~$\Mf^2$.
Although the same effect occurs within the $B_\tau$~mode,
it is of less practical importance in that case,
because the gauge boson masses $M_W$~and~$M_Z$
are of the same order of magnitude as the Higgs mass~$m$,
while the fermion masses are generally much smaller.

\isection{Conclusion}

We have resolved the difficulties associated with
the change of particle content across a domain wall
at a symmetry-breaking phase transition
and obtained scattering solutions for scalar and gauge bosons.
The relationship between the modes $\p2$~and~$B_\l$
provides a precise statement of the intuition that
in a spontaneously broken gauge theory
each apparent Goldstone boson turns into
the longitudinal polarization of a massive gauge boson.

Our results should prove fundamental
to many different calculations involving
the electroweak phase transition in the standard model.
Although the scattering probabilities
for asymptotic particles interacting with a domain wall
are interesting for some purposes,
we expect that the scalar internal modes,
which are capable of describing the gradual change
in particle content across a domain wall,
will prove a more useful tool in actual calculations.

Note added:
We thank M.~Voloshin for bringing to our attention
earlier related work on this subject,
in which the spontaneous breaking of a discrete symmetry
was studied using a quartic potential.
The single bosonic mode present in that situation
obeys the same equations as our $h$~mode.
Polyakov~\cite{polyakov}
found the masses of the bound states of this mode.
Voloshin~\cite{voloshin}
obtained the scattering equation and hypergeometric solutions for this mode,
including the bound-state wavefunctions,
and observed that the transmission probability is identically one.
In the latter work the scattering
of fermions axially coupled to the scalar field was also studied.

\bibliographystyle{plain}

\end{document}